\documentclass[11pt]{article}
\usepackage{graphicx,epsfig,xcolor}
\usepackage[english]{babel}
\usepackage{amssymb,amsfonts,amsmath,bm,bbm,bbold
}
\usepackage{verbatim}
\usepackage{subcaption}

\usepackage{jheppub}

\newcommand{\im}{{i\mkern1mu}}

\newcommand{\arctanh}{\operatorname{arctanh}}
\newcommand{\dd}{\mathrm{d}}

\begin{document}

\title{AdS to dS flows dual to pseudo-Hermitian theories}

\author[a,b]{Jose M. Begines,}
\author[a,b]{Ant\'on F. Faedo,}
\author[a,b]{Carlos Hoyos,}
\author[a,b]{Daniele Musso}
\affiliation[a]{Department of Physics, Universidad de Oviedo, c/ Leopoldo Calvo Sotelo 18, ES-33007 Oviedo, Spain}
\affiliation[b]{Instituto de Ciencias y Tecnolog\'{\i}as Espaciales de Asturias, c/ Independencia 13, ES-33004 Oviedo, Spain}
\emailAdd{jmbsphysics@outlook.es}
\emailAdd{hoyoscarlos@uniovi.es}
\emailAdd{anton.faedo@uniovi.es}
\emailAdd{mussodaniele@uniovi.es}

\abstract{We construct generalizations of ``centaur'' geometries in more than two dimensions as solutions of Einstein gravity coupled to a ghost scalar. The solutions interpolate between asymptotically anti-de Sitter and asymptotically de Sitter spacetimes, with the two asymptotic regions separated by a horizon. A holographic interpretation is that they are dual to pseudo-Hermitian theories obtained from the continuation of real couplings and operators to purely imaginary values. One family of solutions is obtained from a supergravity theory that is a consistent truncation with a known embedding in string theory. We study the thermodynamic and linear stability of the solutions, and determine the uplift of the supergravity solutions to eleven dimensions.}

\maketitle
\flushbottom
\setcounter{page}{2}

\newpage

\section{Introduction}

Understanding quantum gravity in de Sitter spacetime (dS) has been a major challenge in semiclassical approaches and in string theory, to the point that according to some conjectures in the Swampland program, stable or metastable dS geometries cannot exist in a consistent UV-complete theory \cite{Obied:2018sgi}. This is in contrast to Anti-de Sitter (AdS) spacetimes, where the AdS/CFT correspondence has established a precise duality to a UV-complete quantum field theory that one could in principle use to address any of the conceptual issues surrounding quantum gravity. 

Not entering into whether a pure dS spacetime is a \emph{bona fide} vacuum, our own universe is close to being dS, now and during the inflationary era. In the same way that an eternal black hole is useful to better understand black holes in general, we could gain some insight about quantum cosmology by studying theories on dS or almost dS spacetimes. 

A better understanding of dS could be garnered by studying a gravitational background interpolating between an asymptotically AdS spacetime, with a timelike boundary, and an asymptotically dS spacetime, with a future spacelike boundary. Applying the usual AdS/CFT dictionary, one can presumably extract information about dS, by studying observables that probe the asymptotically dS region of the geometry.

Backgrounds interpolating between AdS and dS have been constructed in two dimensions, the so-called ``centaur'' geometries \cite{Anninos:2017hhn,Anninos:2018svg}. In higher dimensions, ``wineglass'' geometries use Euclidean AdS wormholes to set up an initial state for inflation \cite{Betzios:2024oli}. In Lorentzian signature it is possible to glue AdS and dS geometries using a thin wall interface \cite{Alberghi:1999kd,Freivogel:2005qh,Lowe:2010np,Kiritsis:2025ytb}, but there are obstructions to finding regular geometries, at least in theories of Einstein gravity minimally coupled to scalar fields \cite{Kiritsis:2019wyk,Kiritsis:2025ytb}.

A key issue observed in \cite{Kiritsis:2019wyk} is that the null energy condition prevents the existence of regular flows between AdS and dS. This is reminiscent of traversable wormholes: in order for them to exist, one needs to introduce violations of the null energy condition as well, so a way forward may be to do the same for the AdS to dS flows. While this might look like an unphysical thing to do, there is some motivation from the dual field theory point of view that justifies exploring this avenue.

In \cite{Begines:2026mlv} we constructed families of traversable wormholes interpolating between two asymptotically AdS spacetimes that were sourced by ghost scalar fields. These have the `wrong' sign in the kinetic terms and automatically violate the null energy condition. The wormholes have a holographic dual interpretation as states of pseudo-Hermitian field theories obtained by analytic continuation  to imaginary values of real couplings and fields. Although non-unitary, the central charge and conformal dimensions of operators were real and some of the solutions were linearly stable. Furthermore, the wormholes were obtained in analytic continuations of consistent supergravity truncations, where the holographic dictionary is well understood. Related holographic pseudo-Hermitian theories with PT symmetry have been constructed and studied in \cite{Arean:2019pom,Morales-Tejera:2022hyq,Xian:2023zgu,Arean:2024lzz,Arean:2024gks,Arean:2026jie,Maeda:2026awj}.

Following the lessons learned about wormholes, we will look for solutions to Einstein gravity coupled to a ghost scalar interpolating between AdS and dS asymptotic regions, obtained from the analytic continuation to imaginary values of a real ordinary scalar, in such a way that the action remains real. Solutions with ghost scalars of this type have been considered in the past \cite{Bronnikov:2005gm}, but without explicit constructions and outside the context of pseudo-Hermitian holography. Similar analytic extensions of supergravity fields to complex values have been considered in \cite{Bergshoeff:2007cg,Skenderis:2007sm} in the context of the domain wall/cosmology correspondence \cite{Skenderis:2006jq}. 

Applying the holographic dictionary we deduced from the wormhole solutions, the dual field theory would be non-unitary but pseudo-Hermitian, and it is obtained through the analytic continuation of the coupling and operator dual to the ordinary scalar. It should be noted that non-unitarity already made an appearance in several ways in the domain wall/cosmology correspondence \cite{Skenderis:2006fb}, dS solutions of II${}^*$ supergravity theories \cite{Hull:1998vg}, the dS/CFT correspondence put forward in \cite{Witten:2001kn,Strominger:2001pn,McFadden:2009fg}, and in the Kosmic Field Theories \cite{Thavanesan:2025ibm}, the latter also involving an analytic continuation of field theory parameters. This suggests that a holographic dual field theory able to capture dS physics must do away with unitarity. Nevertheless, a pseudo-Hermitian theory could still be a sensible quantum theory \cite{Mostafazadeh:2008pw,Ashida:2020dkc,Bender:2023cem}, which further motivates our approach.

The paper is organized as follows: in Section~\ref{sec:fake} we use a fake supergravity approach to construct a family of theories with analytic AdS to dS flows. In Section~\ref{sec:sugra} we study a supergravity consistent truncation with a real scalar, for which we find numerical AdS to dS flows when the scalar is continued to imaginary values. We also find the uplift to eleven-dimensional supergravity. In Section~\ref{sec:thermo} we study the thermodynamic stability of the solutions, and in Section~\ref{sec:qnms} we address the dynamical stability by computing the low-lying spectrum of linearized fluctuations. In Section~\ref{sec:discuss} we discuss our results and possible future directions. Some technical details are described in the appendices. Specifically, Appendix~\ref{app:numericsSUGRA}
describes the numerical calculations of the supergravity backgrounds; Appendix~\ref{app:holoren} concerns the holographic renormalization and Appendix~\ref{app:qnms} illustrates the numerical computation of the fluctuation (pseudo-)spectra.

\section{Fake supergravity models}\label{sec:fake}

Our goal in this section is to find simple models that admit analytical solutions interpolating between AdS and dS spaces. We will follow the approach of \cite{Kiritsis:2019wyk,Kiritsis:2025ytb,Caceres:2025xzl}, which allows one to find solutions by solving a system of first order equations. The models are a generalization to black brane geometries of the fake supergravity models for Poincar\'e invariant spacetimes \cite{Freedman:2003ax}. 

 The simplest models consist of Einstein gravity minimally coupled to a scalar in $d+1$ dimensions\footnote{We use mostly plus signature for the metric throughout the paper. In our conventions uppercase Latin indices $M,N,\dots$ will label all the directions in the bulk, while Greek indices $\mu,\nu,\dots$ will label the directions parallel to the boundary. When the directions are only spatial we will use lowercase Latin indices $i,j,\dots$.}
\begin{align}\label{eq:action}
S=\frac{1}{2\kappa^2}\int\dd^{d+1}x\ \sqrt{-G}\left[R-\frac12\partial_M\Phi\partial^M\Phi-V(\Phi)\right]\,.
\end{align}
We will look for asymptotically ${\rm AdS}_{d+1}$ solutions and allow the scalar to take real or purely imaginary values. Note that, in the latter case, the kinetic term is real but the sign flips, so the field becomes a ghost. When the potential is even in the scalar field, $V(-\Phi)=V(\Phi)$, the whole action remains real.

The equations of motion following from this action are 
\begin{subequations}\label{eq:eoms}
\begin{align}
R_{MN}-\frac12\partial_M\Phi\partial_N\Phi-\frac{1}{d-1}\,G_{MN}\,V(\Phi)&=0\,,\\[2mm]
\frac{1}{\sqrt{-G}}\partial_M\left(\sqrt{-G}\,\partial^M\Phi\right)-\frac{\dd V(\Phi)}{\dd\Phi}&=0\,.
\end{align}
\end{subequations}
We propose an ansatz for the background metric $G_{MN}=g_{MN}(\rho)$ and scalar $\Phi=\Phi(\rho)$ that takes the form of a black brane solution
\begin{align}\label{eq:metricansatz}
\dd s^2=g_{MN}\, \dd x^M \dd x^N=\frac{\dd\rho^2}{f(\rho)}+e^{2A(\rho)}\left[-f(\rho)\,\dd t^2+\dd x_{d-1}^2\right]\,.
\end{align}
When $f=1$ this is the usual domain wall ansatz and $\rho$ is the domain wall radial coordinate, in which an AdS solution has $A=-\rho/L$ with $L$ the AdS radius and $\rho\to -\infty$ the AdS boundary. Furthermore, if $f=-1$ an equivalent solution $A=\rho/L$ describes the Poincar\'e patch of de Sitter with Hubble constant $H=1/L$ and $\rho$ playing the role of time coordinate and $\rho\to +\infty$ the future boundary. Thus, the additional function $f(\rho)$ acts as a blackening factor that, as extensively discussed in \cite{Kiritsis:2019wyk,Kiritsis:2025ytb}, enables the search for flows between AdS and dS geometries. Notice that, in order for such a flow to exist, the function $f$ has to switch sign, meaning that there must be a value for the radial coordinate $\rho=\rho_h$ at which it vanishes, $f(\rho_h)=0$. This constitutes a horizon. Note also that for a flow that interpolates between AdS and dS, the warp factor grows in both asymptotic regions. Therefore, at some point in the interior the warp factor has a minimum where $A'=0$. We will refer to this point as the ``throat''.

In the metric \eqref{eq:metricansatz}, the equations of motion \eqref{eq:eoms} read  
\begin{subequations}\label{eq:eomsansatz}
\begin{align}
\label{subeq:2.4a}
2(d-1)A''+(\Phi')^2&=0\,,\\[2mm]
\label{eq:eomsansatzf} f''+d\,f'A'&=0\,,\\[2mm]
(d-1)A'\,f'+f\left[d(d-1)(A')^2-\frac12(\Phi')^2\right]+V&=0\,,\\[2mm]
\Phi''+\left(d\,A'+\frac{f'}{f}\right)\Phi'-\frac{1}{f}\frac{\dd V}{\dd \Phi}&=0\,,
\end{align}
\end{subequations}
where prime denotes derivative with respect to $\rho$. These equations can be simplified using the scalar $\Phi$ itself as the radial coordinate. For this, we introduce a superpotential function  
\begin{align}\label{eq:W}
W(\Phi)\equiv-2(d-1)A'\,, \qquad\qquad \Phi'=\frac{\dd W}{\dd\Phi}\,,
\end{align}
such that equation \eqref{subeq:2.4a} is automatically satisfied. The second equation in \eqref{eq:W} can be used to translate derivatives with respect to $\rho$ into derivatives with respect to $\Phi$. In this way the system reduces to \cite{Kiritsis:2019wyk,Kiritsis:2025ytb,Caceres:2025xzl}
\begin{subequations}
\begin{align}\label{eq:eomsWf}
\frac{\dd W}{\dd \Phi}\,\frac{\dd^2f}{\dd \Phi^2}+\left(\frac{\dd^2W}{\dd\Phi^2}-\frac{d}{2(d-1)}W\right)\frac{\dd f}{\dd \Phi}&=0\,,\\[2mm]
\left[\frac{d}{4(d-1)}W^2-\frac12\left(\frac{\dd W}{\dd\Phi}\right)^2\right]f-\frac12\frac{\dd W}{\dd\Phi}\,\frac{\dd f}{\dd\Phi}\,W+V&=0\,. \label{eq:eomsWf2}
\end{align}
\end{subequations}
The second one is the usual relation between potential and superpotential in the domain wall case in which $f=1$ and thus the first one is trivial. Notice that, given a potential $V$, it is in general complicated to find a superpotential satisfying this relation unless there is additional structure like supersymmetry. Taking this into account, we will reverse the logic and propose a sensible superpotential with the desired properties, then solve for the blackening factor $f$ in equation \eqref{eq:eomsWf} and finally use \eqref{eq:eomsWf2} to define the associated potential.

\subsection{A family of simple examples}

We will now take advantage of the fake supergravity formalism to produce a family of simple analytic solutions that include AdS to dS flows. Supergravity theories usually have scalar potentials depending on exponentials of the field. Using this as inspiration, a possible superpotential is  
\begin{align}
W(\Phi)=\frac{2(d-1)}{L}\,\cosh\left(\frac{\sqrt{d}}{2\sqrt{d-1}}\,\Phi\right)\,.
\end{align}
We have tuned the parameters so that we can easily solve the equations \eqref{eq:eomsWf} and \eqref{eq:eomsWf2} and find flows with the desired properties. 

\subsubsection{Real flows}

For a real flow $\Phi=\phi$, the blackening function associated to the superpotential is 
\begin{align}\label{eq:f}
f=1+f_0\left[\cosh\left(\frac{\sqrt{d}}{2\sqrt{d-1}}\,\phi\right)-1\right]\,,
\end{align} 
where we have fixed an integration constant so that $f\to1$ when $\phi\to0$. From the equation in \eqref{eq:eomsWf2} one can extract the potential, which for a real scalar reads
\begin{align}\label{eq:Vbottomup}
V=-\frac{d(d-1)}{L^2}\left[f_0\cosh\left(\frac{\sqrt{d}}{2\sqrt{d-1}}\,\phi\right)+\frac{(1-f_0)}{4}\left(3+\cosh\left(\frac{\sqrt{d}}{\sqrt{d-1}}\,\phi\right)\right)\right]\,.
\end{align}
The only extremum of the superpotential at a real value of the scalar is at $\phi_0=0$, which is shared with the potential. Expanding the potential at the extremum
\begin{align}
V=-\frac{d(d-1)}{L^2}-\frac{d^2}{8L^2}\phi^2+\cdots\,,
\end{align} 
so that this corresponds to an AdS solution with radius $L$. The mass of the scalar saturates the Breitenlohner--Freedman (BF) bound $m^2_{\text{\tiny AdS}}L^2=-\frac{d^2}{4}$, so the dual operator has conformal dimension $\Delta=\frac{d}{2}$.

The potential has additional extrema at 
\begin{align}
\tilde{\phi}_0^\pm=\frac{2\sqrt{d-1}}{\sqrt{d}}\log\left(\frac{f_0\pm\sqrt{2f_0-1}}{f_0-1}\right)\,,\quad\quad f_0>1\,.
\end{align}
The values $\tilde{\phi}_0^\pm$ are real for $f_0>1$ and correspond to minima of the potential. On the other hand, in the range $\frac12<f_0<1$ both are complex and for $f_0\leq \frac{1}{2}$ they become purely imaginary. The form of the potential for real values of the scalar and different values of $f_0$ is depicted in Figure~\ref{fig:V}(a).

\subsubsection{Purely imaginary flows}

Extending $\Phi$ to complex values, the superpotential is periodic along the imaginary axis
\begin{equation}
   W\left(\Phi+\frac{4\sqrt{d-1}}{\sqrt{d}}\pi\im\right)=W(\Phi)\,.
\end{equation}
This periodicity extends to the blackening factor and the potential. For a purely imaginary background $\Phi=\im\eta$, with $\eta\in \mathbb{R}$, the blackening factor is
\begin{align}\label{eq:imf}
f=1+f_0\left[\cos\left(\frac{\sqrt{d}}{2\sqrt{d-1}}\,\eta\right)-1\right]\,.
\end{align} 
The function $f$ remains real for real values of $f_0$.
The potential along the imaginary direction is
\begin{align}\label{eq:Vbottomup2}
V=-\frac{d(d-1)}{L^2}\left[f_0\cos\left(\frac{\sqrt{d}}{2\sqrt{d-1}}\,\eta\right)+\frac{(1-f_0)}{4}\left(3+\cos\left(\frac{\sqrt{d}}{\sqrt{d-1}}\,\eta\right)\right)\right]\,.
\end{align}
Thus, there are infinitely many copies of the $\Phi=0, \tilde\phi_0^\pm$ extrema at values separated by an integer multiple of the imaginary period. Alternatively, one could think of extrema separated by a period as being equivalent, so field space is compactified along the imaginary direction. 

Along the imaginary direction the superpotential has additional extrema at 
\begin{align}\label{eq:Wext}
\eta_0=\frac{2\sqrt{d-1}}{\sqrt{d}}\,\pi\,,
\end{align}
and values related by a shift by the imaginary period.
Notice that extrema of the superpotential are also extrema of the potential when $f$ is constant. Expanding around \eqref{eq:Wext}
\begin{align}
V=(1-2f_0)\left[-\frac{d(d-1)}{L^2}+\frac{d^2}{8L^2}(\eta-\eta_0)^2\right]+\cdots\,.
\end{align} 
The nature of the extrema depends on the value of $f_0$, being AdS minima for $f_0<\frac12$ and dS maxima if $f_0>\frac12$. At the crossover value $f_0=\frac12$ the potential vanishes and the solution is Ricci flat. The radius of the AdS solutions at these extrema is $\tilde L=L/\sqrt{1-2f_0}$. When the extrema are dS, the Hubble constant is $H=\sqrt{2f_0-1}/L$. Taking into account the reversed kinetic term of the scalar along the imaginary direction, it saturates the BF bound at the AdS minima and has a positive mass at the dS maxima $m_{\text{\tiny dS}}^2=\frac{d^2}{4}H^2$.

\begin{figure*}[t!]
    \centering
    \begin{subfigure}[t]{0.48\textwidth}
    \centering
    \includegraphics[width=\linewidth]{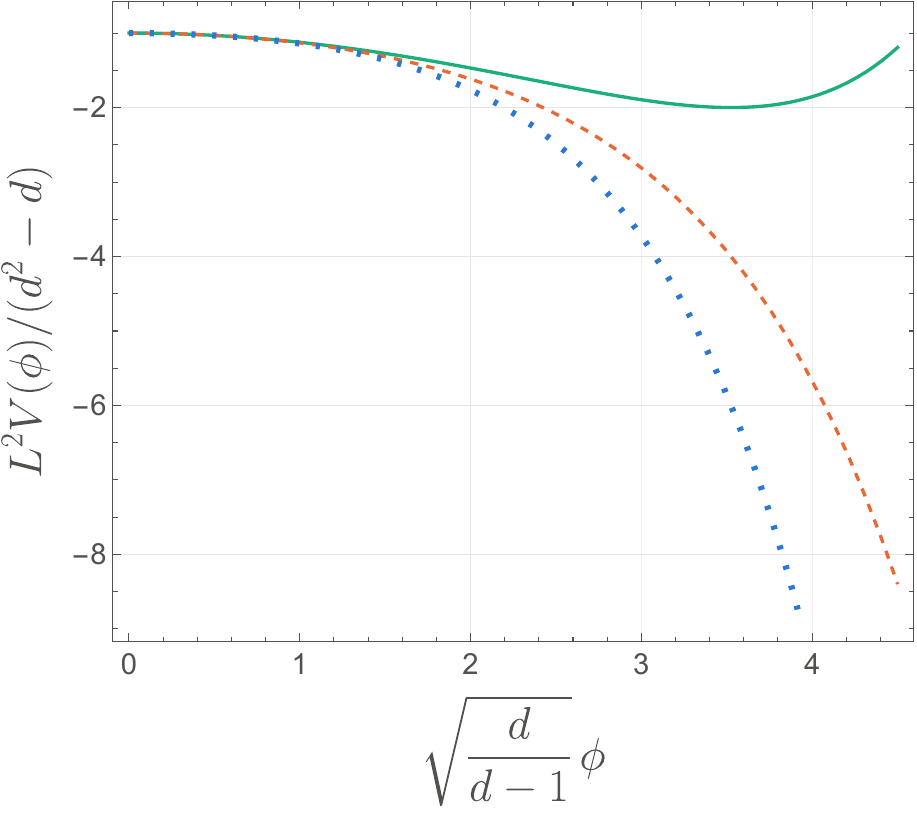}
    \caption{Potential for real values of $\Phi$.}
    \end{subfigure}
    \quad
    \begin{subfigure}[t]{0.48\textwidth}
    \centering
    \includegraphics[width=\linewidth]{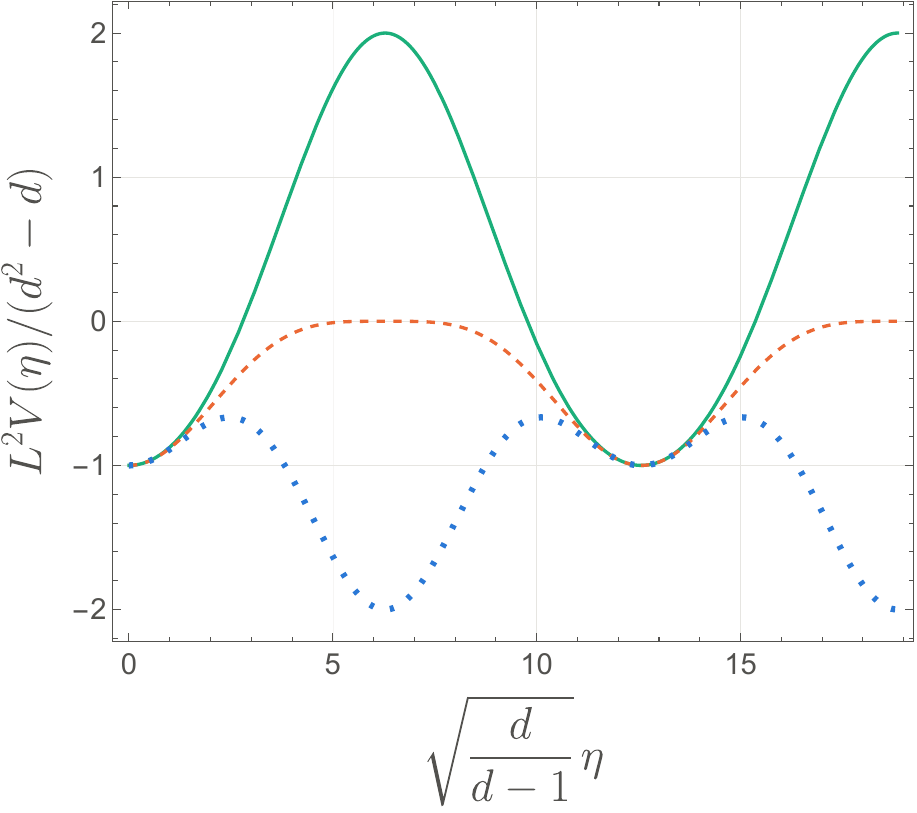}
\caption{Potential for imaginary values of $\Phi$.}
    \end{subfigure}
    \caption{Potential $\frac{L^2V}{d(d-1)}$ as a function of the scalar $\sqrt{\frac{d}{d-1}}\,\Phi$ for $f_0=\frac32$ (green, continuous), $f_0=\frac12$ (dashed, red) and $f_0=-\frac12$ (dotted, blue). It can be seen that the real minimum is lost when $f_0\le1$.}\label{fig:V}
\end{figure*}

As mentioned above, the $\tilde{\phi}_0^\pm$ extrema become purely imaginary for $f_0\leq \frac{1}{2}$. Exactly at $f_0=\frac12$ they merge with the extrema \eqref{eq:Wext}. When $f_0<\frac12$ they are purely imaginary and, furthermore, $\tilde{\phi}_0^\pm$ become the closest ones to the real axis. They merge with the extremum at $\Phi=0$ in the limit $f_0\to-\infty$. The potential along imaginary values of the scalar is depicted in Figure~\ref{fig:V}(b).

\subsubsection{AdS to dS flows}

Since the potential is still real for purely imaginary values of the scalar, the metric would also be real if $\Phi$ explores the imaginary axis from the AdS extremum at $\Phi=0$ to an imaginary extremum which, interestingly, would be dS for $f_0>\frac12$. To describe this RG flow, one solves the second equation in \eqref{eq:W}, which, for $\Phi=\im\eta$, gives
\begin{equation}\label{eq:scalarsol}
\eta=\frac{4\sqrt{d-1}}{\sqrt{d}}\arctan\left[ \exp\left(\frac{d(\rho-\rho_0)}{2L}\right)\right]\,.
\end{equation}
This clearly interpolates between $\eta=0$ at $\rho\to-\infty$ and $\eta=\eta_0$ (cf.~\eqref{eq:Wext}) at $\rho\to +\infty$. Note that for $f_0<\frac{1}{2}$, like the blue dotted curve in Figure~\ref{fig:V}(b), the solution interpolates between two minima and skips the maximum in between. This corresponds to wormholes interpolating between two asymptotically AdS regions, similar to the ones constructed in \cite{Begines:2026mlv}. In this case the two AdS regions have different radii in general, except for $f_0=0$, when $\tilde L=L$.

Plugging the scalar solution \eqref{eq:scalarsol} into the first equation in \eqref{eq:W} we get the warp factor 
\begin{equation}\label{eq:eA}
    e^A=\left[2e^{-\frac{d\rho_0}{2L}}\cosh\left(\frac{d(\rho-\rho_0)}{2L}\right)\right]^{2/d}\,,
\end{equation}
where we have fixed an integration constant in such a way that $A\to-\frac{\rho}{L}$ when $\rho\to-\infty$ so that we recover an AdS solution with radius $L$ in that limit. Finally the blackening factor for this solution is simply 
\begin{align}\label{eq:f2}
f=1-f_0\left[1+\tanh\left(\frac{d(\rho-\rho_0)}{2L}\right)\right]\,,
\end{align}
which interpolates between $f\to1$ and $f\to1-2f_0$. For $f_0>\frac{1}{2}$, this function vanishes at
\begin{equation}\label{eq:rhoh}
    \rho=\rho_h\equiv \rho_0+\frac{2L}{d}\arctanh\left(\frac{1}{f_0}-1\right)\,.
\end{equation}
This is the position of the horizon separating the AdS and dS regions. A crucial point is that, since both $A$ and $f$ are real, the entire metric \eqref{eq:metricansatz} is also real despite the scalar being imaginary. 

The solution with $f_0=\frac{1}{2}$ has a special behavior. In that case the potential simplifies to  
\begin{align}
V=-\frac{d(d-1)}{L^2}\left[\cos\left(\frac{\sqrt{d}}{4\sqrt{d-1}}\,\eta\right)\right]^4
\end{align}
so that it vanishes at $\eta=\eta_0$. This means that in the limit $\rho\to+\infty$ the metric must be Ricci flat. Indeed, it takes the form 
\begin{align}
\dd s^2\sim e^{-\frac{2\rho_0}{L}}\,e^{-(d-2)\frac{(\rho-\rho_0)}{L}}\dd t^2+e^{-\frac{4\rho_0}{L}}\,e^{\frac{2\rho}{L}}\dd x_{d-1}^2+e^{\frac{d(\rho-\rho_0)}{L}}\,\dd \rho^2\,.
\end{align}
Changing the radial coordinate according to $e^{\frac{d}{2L}(\rho-\rho_0)}\,\dd \rho=\dd z$ and rescaling the Minkowski coordinates the metric becomes  
\begin{align}
\dd s^2\sim -\left(\frac{z}{L}\right)^{-\frac{2(d-2)}{d}}\dd t^2+\left(\frac{z}{L}\right)^{\frac{4}{d}}\dd x_{d-1}^2+\dd z^2\,.
\end{align}
This is a Kasner-type metric in the radial variable $z$ instead of time.

\section{A supergravity example}\label{sec:sugra}

A simple model with a potential that can interpolate between AdS and dS geometries can be found in \cite{Gauntlett:2009bh}. The four-dimensional supergravity action is\footnote{We have reinstated the AdS radius $L$ with respect to $L=1/2$ in \cite{Gauntlett:2009bh}.} 
\begin{align}
S=\frac{1}{2\kappa^2}\int\dd^{4}x\sqrt{-G}\left[R-\frac{3}{2(1-h^2)^2}\partial_Mh\partial^Mh+\frac{6}{(1-h^2)^{1/2}L^2}\right]\,.
\end{align}
To canonically normalize the scalar we define a new field 
\begin{align}
h=\tanh\left(\frac{\Phi}{\sqrt{3}}\right)\,,
\end{align}
in terms of which the action takes the form
\begin{align}\label{eq:Ssugra}
S=\frac{1}{2\kappa^2}\int\dd^{4}x\sqrt{-G}\left[R-\frac{1}{2}\partial_M\Phi\partial^M\Phi+\frac{6}{L^2}\cosh\left(\frac{\Phi}{\sqrt{3}}\right)\right]\,.
\end{align}
The scalar potential has a critical point at $\Phi=0$, where it can be expanded as
\begin{equation}
    V(\Phi)=-\frac{6}{L^2}-\frac{1}{L^2}\Phi^2+\cdots\,.
\end{equation}
Since the potential has a negative maximum, the critical point corresponds to an AdS space with radius $L$. From the quadratic term of the potential we can read the mass of the scalar field around this point $m_{\text{\tiny AdS}}^2L^2=-2$. Following the usual AdS/CFT dictionary, the scalar field is dual to a scalar operator of conformal dimension $\Delta=2$ (or $\Delta=1$ in alternative quantization).

 When the scalar field is extended to imaginary values, $\Phi=\im\eta$, the action \eqref{eq:Ssugra} remains real. The kinetic term of the scalar in \eqref{eq:Ssugra} has its sign reversed, so it becomes a ghost, with potential
\begin{equation}
    V(\eta)=-\frac{6}{L^2}\cos\left(\frac{\eta}{\sqrt{3}}\right)\,.\label{eq:VsugraIm}
\end{equation}
The potential is periodic in $\eta$, with period $2\sqrt{3}\pi$. Critical points in the $[0,2\sqrt{3}\pi)$ interval lie at $\eta=0$ and $\sqrt{3}\pi$. Expanding around $\eta=0$,
\begin{equation}\label{eq:SprGravPot}
    V(\eta)=-\frac{6}{L^2}+\frac{1}{L^2}\eta^2+\cdots\,.
\end{equation}
Since the potential has a negative minimum, the geometry at the critical point is AdS with radius $L$. Expanding to linear order around this point, the equation of motion for the ghost field is the same as the one of an ordinary scalar with mass $m_{\text{\tiny AdS}}^2 L^2=-2$. Therefore, the ghost is dual to the analytic extension of the operator dual to $\Phi$.

On the other hand, if we expand around $\eta=\sqrt{3}\pi$, the potential
\begin{equation}
    V(\eta)=+\frac{6}{L^2}-\frac{1}{L^2}(\eta-\sqrt{3}\pi)^2+\cdots\,.
\end{equation}
At this critical point the potential has a positive maximum, so the geometry is dS with Hubble constant $H=1/L$. The mass of the ghost field at this point is positive $m_{\text{\tiny dS}}^2L^2=2$.

Other critical points related to the two just described by a shift by multiples of the period have the same characteristics. We will look for solutions interpolating between neighboring AdS minima and dS maxima of the potential.

\subsection{AdS to dS flows}

We follow the same conventions as for the fake supergravity models
and use the following domain-wall ansatz for the metric and the ghost scalar field
\begin{equation}\label{eq:metricSUGRA}
    \dd s^2=\frac{\dd \rho^2}{f(\rho)}+e^{2A(\rho)}\left[-f(\rho)\,\dd t^2+\dd x_2^2\right],\qquad \eta=\eta(\rho)\,.
\end{equation}
We will impose the conditions that the geometry interpolates between AdS at $\rho\to -\infty$ and dS at $\rho\to +\infty$, that is,     
\begin{subequations}
\begin{align}
    &\lim_{\rho\to-\infty} \eta=0\,,\qquad\quad \lim_{\rho\to-\infty} f=1\,,\qquad\quad A\underset{\rho\to -\infty}{\sim} -\frac{\rho}{L}\,, \label{eq:asympcond} \\  
    &\lim_{\rho\to+\infty} \eta=\sqrt{3}\pi\,,\qquad\quad \lim_{\rho\to+\infty} f=-\frac{1}{a_\infty^2}\,,\qquad\quad A\underset{\rho\to +\infty}{\sim} a_\infty\frac{\rho}{L}\,.
\end{align}
\end{subequations}
Note that $\rho$ is both the holographic radial coordinate in the AdS region and the time coordinate in the dS region. The constant $a_\infty$ is determined after solving the equations numerically. It can be removed by changing the coordinate $\rho$ to the proper distance/time. Since $f$ has opposite signs in the AdS and dS regions, there must be a value $\rho=\rho_h$ such that $f(\rho_h)=0$, which corresponds to a horizon. Therefore, the metric describes an asymptotically AdS black brane geometry. Inside the horizon, instead of the usual type of spacetime ending at a future singularity, the geometry is smooth and ends at an asymptotically dS future boundary.

The Einstein and scalar equations of motion are the same as \eqref{eq:eomsansatz} with $d=3$. Imposing the boundary conditions \eqref{eq:asympcond}, the asymptotic expansions of the background functions are
\begin{align}\label{eq:rhoexpansions}
    f&=1+f_3 \,e^{\frac{3\rho}{L}}+O\left(e^{\frac{5\rho}{L}}\right),\qquad\quad \eta=\eta_1\, e^{\frac{\rho}{L}}+\eta_2\, e^{\frac{2\rho}{L}}+O\left(e^{\frac{3\rho}{L}}\right)\,,\nonumber\\[2mm]
    e^{A}&=e^{-\frac{\rho}{L}}\left[ 1+ \frac{\eta_1^2}{16} \,e^{\frac{2\rho}{L}}+\frac{\eta_1\eta_2}{9} \,e^{\frac{3\rho}{L}}+O\left(e^{\frac{4\rho}{L}}\right)\right]\,,
\end{align}
with all the subleading terms fixed by the values of the coefficients $f_3$, $\eta_1$ and $\eta_2$. As we will see, the parameters $\eta_1$ and $\eta_2$ determine respectively the source and expectation value of the operator dual to the scalar $\eta$.
Together with $f_3$, they also determine the expectation values of the energy density and pressure.

\begin{figure}[t]
    \centering
    \begin{subfigure}[t]{0.49\textwidth}
        \centering
        \includegraphics[width=\linewidth]{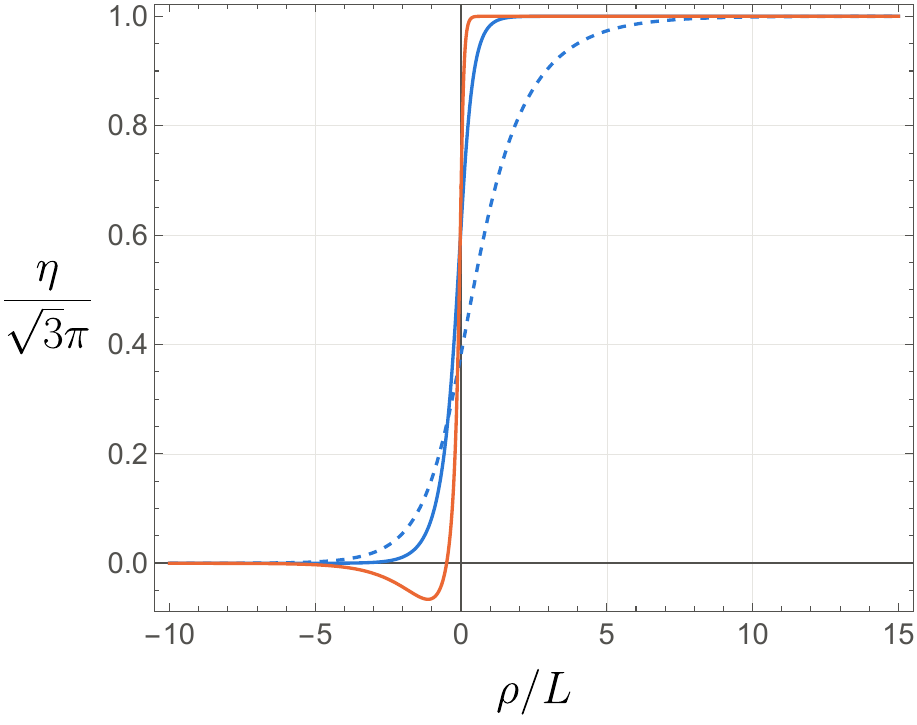}
        \caption{}
        \label{fig:flow-eta}
    \end{subfigure}\hfill
    \begin{subfigure}[t]{0.47\textwidth}
        \centering
        \includegraphics[width=\linewidth]{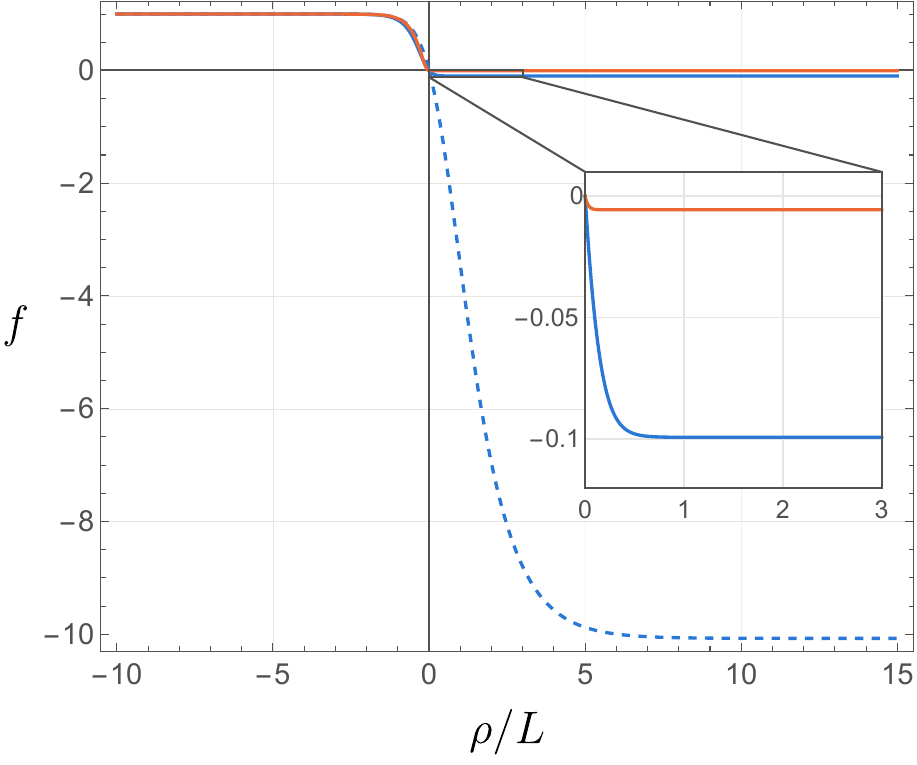}
        \caption{}
        \label{fig:flow-f}
    \end{subfigure}\\[2mm]
    \begin{subfigure}[t]{0.48\textwidth}
        \centering
        \includegraphics[width=\linewidth]{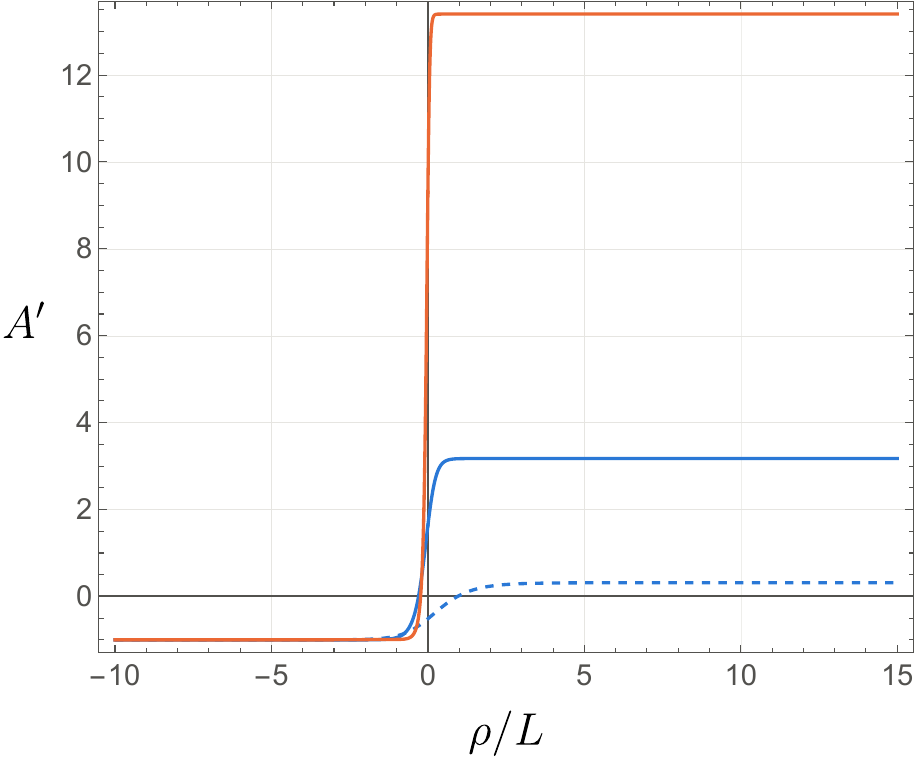}
        \caption{}
        \label{fig:flow-Aprime}
    \end{subfigure}
\caption{Background solutions for the supergravity example with action \eqref{eq:Ssugra} where we represent the ghost field $\eta$ (a), the blackening factor $f$ (b) and the derivative of the warp factor $A'$ (c) for three solutions. The blue lines are solutions with no bounces, with the throat behind the horizon (dashed) or outside (solid). The red solid lines correspond to a solution with one bounce.
}\label{fig:solSUGRA}
\end{figure}

We solve the equations numerically by shooting from the horizon towards the AdS and the dS boundaries.
Details of the numerical calculation are given in Appendix~\ref{app:numericsSUGRA}. Plots for some of the solutions are shown in Figure~\ref{fig:solSUGRA}.
 
We find different families of solutions, characterized by the number of times the radial derivative of the scalar field changes sign in the region between the AdS boundary and the horizon. These are ``bounces'', similar to the ones described in a different context in \cite{Kiritsis:2016kog, Gursoy:2018umf}. Each bounce corresponds to an oscillation around the AdS minimum of the potential before the scalar settles there at the asymptotic boundary. 
It turns out that, close to the dS maximum of the scalar potential, there is a critical value of the scalar at the horizon $\eta_c\approx 0.7311 \sqrt{3}\pi$ where the number of bounces increases without bound. Beyond the critical value there is a runaway behavior, such that the solutions do not have an asymptotic AdS boundary. This limits solutions with a holographic dual interpretation to those where the value of the scalar at the horizon is below the critical value. All the solutions with bounces have the throat outside the horizon. Exploring the region inside the horizon, it turns out that an analogous behavior is present when the value of the scalar at the horizon comes closer to the AdS minimum of the scalar potential. There are bounces of the scalar inside the horizon whose number increases without bound as one approaches a critical value $\eta_c'=\sqrt{3}\pi-\eta_c\approx 0.2689\sqrt{3}\pi$, and beyond which there are no solutions approaching dS in the asymptotic future.

\subsection{UV completion of the supergravity model}\label{sec:uplift}

The action \eqref{eq:Ssugra} can be obtained from a consistent truncation of eleven-dimensional supergravity on an arbitrary seven-dimensional Sasaki--Einstein manifold \cite{Gauntlett:2009bh}.\footnote{The very same action can be obtained from the truncation to the four-dimensional STU model in \cite{Cvetic:1999xp}, but the corresponding uplifted metric is not real for imaginary values of the scalar.} This includes the seven-sphere, so among the possible dual gauge theories there is ABJM. The truncation  contains both a charged and a neutral scalar together with a vector. It is consistent to switch off the charged scalar and the vector simultaneously. 

In terms of the canonically normalized, four-dimensional scalar, the metric takes the form 
\begin{align}\label{eq:11dmetric}
\dd s_{11}^2=\sigma \cosh\left(\frac{\Phi}{\sqrt{3}}\right)\dd s_4^2+4L^2\dd s_6^2\left({\rm KE}\right)+4L^2\cosh^{-2}\left(\frac{\Phi}{\sqrt{3}}\right)\eta^2_{\text{\tiny{SE}}}\,,
\end{align}
with $\dd s_6^2\left({\rm KE}\right)$ the metric of the K\"ahler--Einstein base, normalized so that the curvature is $R_{\text{\tiny KE}}=48$, and $\eta_{\text{\tiny{SE}}}$ the fiber over it. The parameter $\sigma^2=1$ will be important after the analytic extension, it allows one to keep the same signature throughout the solution.

This metric is supported by the four-form flux 
\begin{align}\label{eq:F4flux}
F_4=\frac{3}{L}\cosh\left(\frac{\Phi}{\sqrt{3}}\right){\rm vol}_4+8L^3 \dd\left[\tanh\left(\frac{\Phi}{\sqrt{3}}\right) J\wedge\eta_{\text{\tiny{SE}}}\right]\,,
\end{align}
with $J$ the K\"ahler form on the base, satisfying $\dd \eta_{\text{\tiny{SE}}}=2J$. For more details on this, see \cite{Gauntlett:2009bh}.

The extension to imaginary values of the scalar, $\Phi=\im\eta$, results in the following metric and fluxes
\begin{subequations}
    \begin{align}
\dd s_{11}^2&=\left|\cos\left(\frac{\eta}{\sqrt{3}}\right)\right|\dd s_4^2+4L^2\dd s_6^2\left({\rm KE}\right)+4L^2\cos^{-2}\left(\frac{\eta}{\sqrt{3}}\right)\eta^2_{\text{\tiny{SE}}}\,,\label{eq:11dmetric2}\\[2mm]
F_4&=\frac{3}{L}\cos\left(\frac{\eta}{\sqrt{3}}\right){\rm vol}_4+8\im L^3 \dd\left[\tan\left(\frac{\eta}{\sqrt{3}}\right) J\wedge\eta_{\text{\tiny{SE}}}\right]\,.   \label{eq:F4flux2}
    \end{align}
\end{subequations}
We have fixed $\sigma$ in such a way that the signature of the metric does not change.
We observe that the eleven-dimensional metric remains real, while the flux becomes purely imaginary along the internal space.

 The uplift to eleven dimensions uncovers a problem, since the metric \eqref{eq:11dmetric2} and flux 
\eqref{eq:F4flux2} become singular in between the AdS and dS points, at $\eta=\frac{\sqrt{3}}{2}\pi$. 
We encounter the curious situation where both the metric and the scalar field are perfectly regular in the reduced theory, but the geometry is singular when uplifted to higher dimensions. Note also that the factor multiplying the four-dimensional metric in \eqref{eq:11dmetric2} has a discontinuous derivative at the point where it vanishes, so it requires the introduction of a codimension-one thin brane localized in the radial direction and gluing the two parts of the geometry. The location of the singularity is consistent with the imaginary distance conjecture \cite{Maldacena:2026jqd, DiUbaldo:2026rly, Etheredge:2026rio}  extended to massive fields \cite{Begines:2026mlv}.

Several observables, proportional to areas of hypersurfaces embedded in the geometry, are blind to the singularity, due to a cancellation of singular and vanishing factors in the eleven-dimensional metric. These include the black hole and entanglement entropies, and the action of M2 branes wrapping the $\eta_{\text{\tiny{SE}}}$ fiber and extended along one spatial direction in the four-dimensional non-compact subspace. Note that in the reduction to type IIA along the fiber, the M2 becomes a fundamental string extended along the non-compact directions. 

\section{Thermodynamics}\label{sec:thermo}

The AdS to dS flows have a horizon separating the two asymptotic regions. From the AdS region point of view, this is a (non-extremal) black brane geometry corresponding in the field theory dual to a thermal state. One can extract thermodynamic properties applying the usual holographic dictionary.

The temperature and entropy density in the dual field theory are derived from the surface gravity (or regularity of the Euclidean solution) and area of the horizon in Planck units, respectively\footnote{Note that $f'(\rho_h)<0$.}
\begin{equation}\label{eq:Ts}
    T=\frac{|f'(\rho_h)|}{4\pi}e^{A(\rho_h)}\,,\qquad\qquad s=\frac{2\pi}{\kappa^2}e^{(d-1)A(\rho_h)}\,.
\end{equation}
Other thermodynamic quantities can be computed from the expectation value of the energy momentum tensor (energy density and pressure) and the on-shell gravity action (free energy density). This requires using holographic renormalization; details are given in Appendix~\ref{app:holoren}.

\subsection{Fake supergravity models}

The asymptotic expansion of the ghost scalar field at the AdS boundary $\rho\to -\infty$ is
\begin{align}
\eta=4\sqrt{\frac{d-1}{d}}\,e^{\frac{d}{2}\frac{(\rho-\rho_0)}{L}}+\dots\,.
\end{align}
This is the normalizable solution, which maps to the expectation value of the dual operator. Since the leading non-normalizable terms $\sim \rho\,  e^{\frac{d}{2}\frac{\rho}{L}}$ are absent, the source for the operator vanishes in the dual field theory. In particular, this means that conformal invariance is only broken spontaneously.

Combining \eqref{eq:eA}, \eqref{eq:f2}, and \eqref{eq:rhoh} in \eqref{eq:Ts} one obtains the following analytic expressions for the temperature and entropy density
\begin{align}
T=\frac{d}{4\pi L}\,(2f_0)^{\frac{2-d}{d}}\,\left(2f_0-1\right)^{\frac{d-1}{d}}\,e^{-\frac{\rho_0}{L}}\,,\qquad s=\frac{2\pi}{\kappa^2}\,(2f_0)^{\frac{2(d-1)}{d}}\,(2f_0-1)^{\frac{1-d}{d}}\,e^{-\frac{(d-1)\rho_0}{L}}
\end{align} 
Solving for $\rho_0$, the entropy density in terms of the temperature is
\begin{equation}
    s=d\left( \frac{4\pi}{d}\right)^{d}\frac{L^{d-1}}{2\kappa^2}\left(\frac{2f_0}{2f_0-1} \right)^{d-1} T^{d-1}\,.
\end{equation}
The expected behavior in a conformal theory, $s\sim T^{d-1}$, is verified. Note that the specific heat $c_V=T\frac{\partial s}{\partial T}>0$, so these solutions are locally thermodynamically stable.

\begin{figure}[t]
\begin{center}
\includegraphics[width=0.60\textwidth]{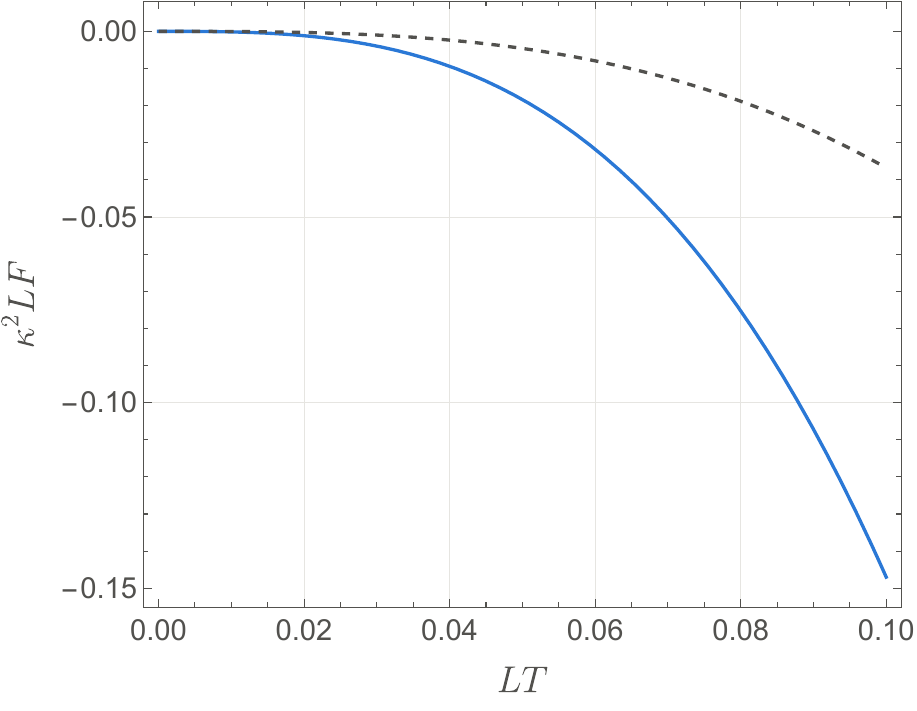}
\caption{Free energy as a function of the temperature for the AdS black hole (dashed line) and for the AdS to dS solution (continuous line) with $f_0=1$ . We see that the latter is smaller for $T>0$, so it dominates thermodynamically at every temperature.}
\label{fig:Free}
\end{center}
\end{figure}

In order to determine their global thermodynamic stability we must compare with the state with zero expectation value for the scalar operator, which exists for every choice of the potential \eqref{eq:Vbottomup}, \emph{i.e.}, for any value of $f_0$. The dual background is just the AdS black brane solution, which in these coordinates reads
\begin{equation}
    e^{A(\rho)}=e^{-\frac{\rho}{L}},\qquad f(\rho)=1-e^{\frac{d(\rho-\rho_h)}{L}},\qquad \eta=0\,.
\end{equation}
Introducing these expressions in \eqref{eq:Ts}, the temperature and entropy density $s_{\text{\tiny AdS}}$ of the black brane are
\begin{equation}
    T=\frac{d}{4\pi L}e^{-\frac{\rho_h}{L}},\ \quad s_{\text{\tiny AdS}}=\frac{2\pi}{\kappa^2}e^{-\frac{(d-1)\rho_h}{L}}\,\ \qquad \Rightarrow \qquad \ s_{\text{\tiny AdS}}=d\left( \frac{4\pi}{d}\right)^{d}\frac{L^{d-1}}{2\kappa^2} T^{d-1}\,.
\end{equation}
From the entropy density we can compute the free energy density, which determines which state is favored thermodynamically, for each value of $f_0$. The free energy density equals minus the pressure, and can be computed integrating the entropy density with respect to the temperature\footnote{The result coincides with the expressions obtained from holographic renormalization in Appendix~\ref{app:holoren}.}
\begin{equation}
    F=-p=-\left( \frac{4\pi}{d}\right)^{d}\frac{L^{d-1}}{2\kappa^2}\left(\frac{2f_0}{2f_0-1} \right)^{d-1} T^d\,,\qquad F_{\text{\tiny AdS}}=-\left( \frac{4\pi}{d}\right)^{d}\frac{L^{d-1}}{2\kappa^2} T^d,
\end{equation}
where $F_{\text{\tiny AdS}}$ denotes the free energy of the AdS black brane. Since $F<F_{\text{\tiny AdS}}$ for $f_0>\frac{1}{2}$, the solutions interpolating between AdS and dS regions would be favored thermodynamically according to the usual criterion (see Figure~\ref{fig:Free}).

\subsection{Supergravity example}

In this case the solution is numeric so we resort to the holographic renormalization results in Appendix~\ref{app:holoren}. First, there is no superpotential that reproduces our solution, so we fix the counterterm as 
\begin{align}
W(\eta)=\frac {4} {L}-\frac{\eta^2}{2L}\,.
\end{align}
This cancels the divergences of the bulk on-shell action and reproduces the AdS black hole results discussed above. Substituting this, together with the expansion \eqref{eq:rhoexpansions}, in the expression for the internal energy density and pressure, eqs.~\eqref{eq:intenergy} and \eqref{eq:pressure}, one obtains 
\begin{align}
\varepsilon=\frac{\eta_1\eta_2-6f_3}{6L\kappa^2}\,,\qquad\qquad p=-\frac{\eta_1\eta_2+3f_3}{6L\kappa^2}\,.
\end{align}
The energy-momentum tensor is not traceless due to the explicit breaking of conformal symmetry in the form of a source for the operator dual to the scalar. In particular, its value is 
\begin{align}
    \langle T^\mu{}_\mu \rangle=-\varepsilon+2p=-\frac{\eta_1\eta_2}{2L\kappa^2}\,.
\end{align}
The free energy density is computed in a similar way from \eqref{eq:free}, taking into account that there is a contribution from the horizon. In terms of the parameters of the expansion it reads
\begin{align}
F=\frac{\eta_1\eta_2-6f_3}{6L\kappa^2}-\frac{e^{3A(\rho_h)}\left|f'(\rho_h)\right|}{2\kappa^2}\,.
\end{align}
The first term can be recognized as the internal energy density, while the second one is just the product of the entropy density and temperature. It is thus verified $F=\varepsilon-sT$ as expected. 

It is possible to write the free energy entirely in terms of the coefficients of the expansion close to the boundary by exploiting the conserved quantity \eqref{eq:conscharge}. Evaluating it at the boundary and at the horizon one gets the following relation between parameters 
\begin{align}
Q(\rho_h)=Q(-\infty)\qquad\Rightarrow\qquad f'(\rho_h)e^{3A(\rho_h)}=\frac{3f_3}{L}\,.
\end{align}
The free energy density takes the final, simple form 
\begin{align}
 F=\frac{\eta_1\eta_2+3f_3}{6L\kappa^2}
\end{align}
in which it is explicit that $F=-p$ and therefore $\varepsilon+p=sT$. 

\begin{figure}
    \centering
    \includegraphics[width=0.48\linewidth]{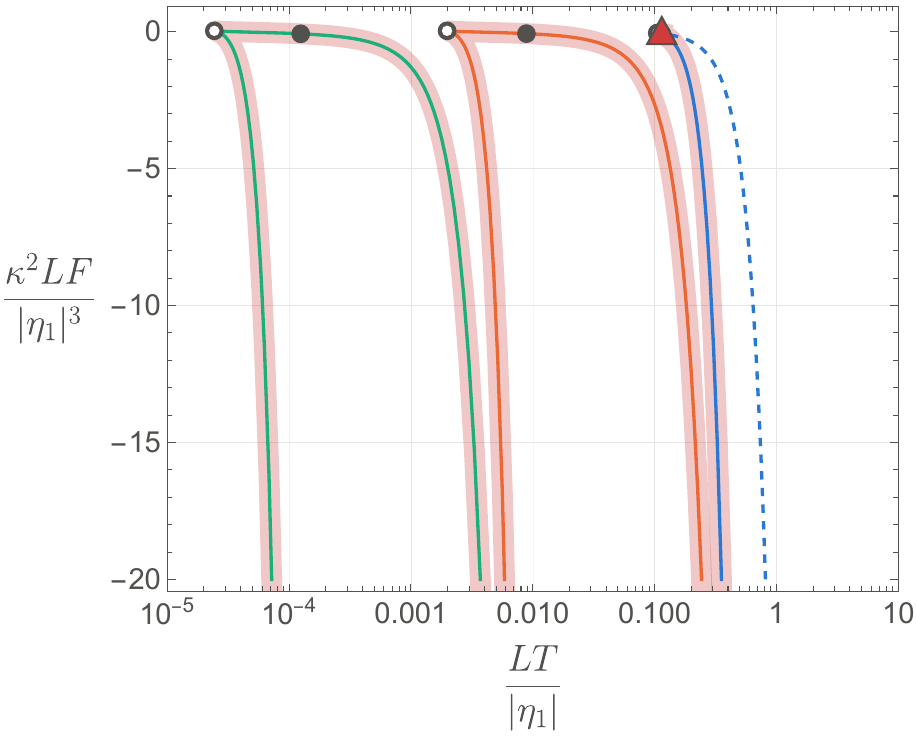}
    \includegraphics[width=0.50\linewidth]{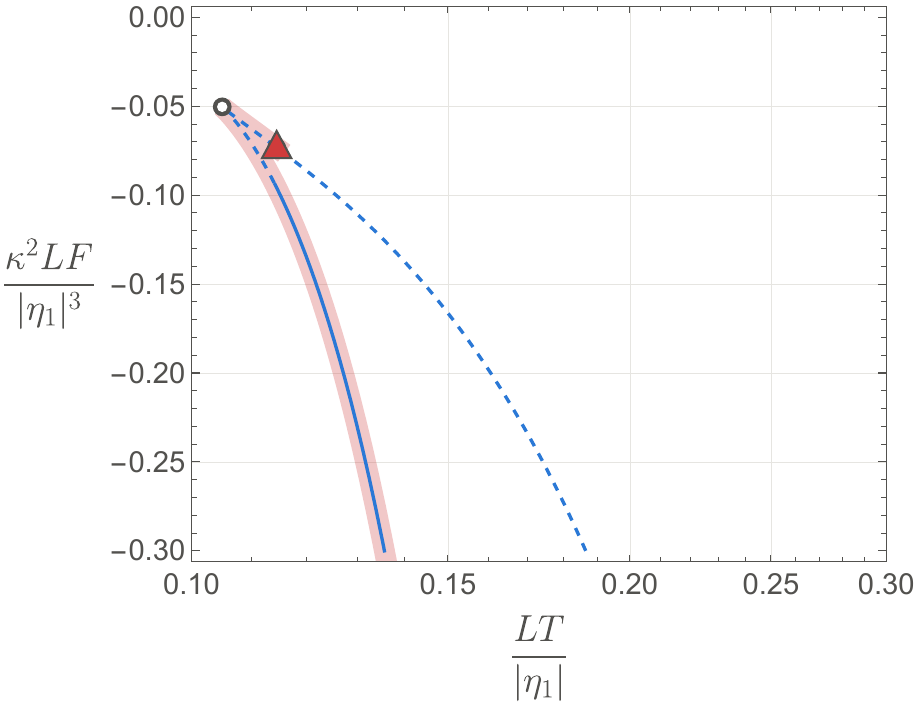}\\[2mm]
    \includegraphics[width=0.49\linewidth]{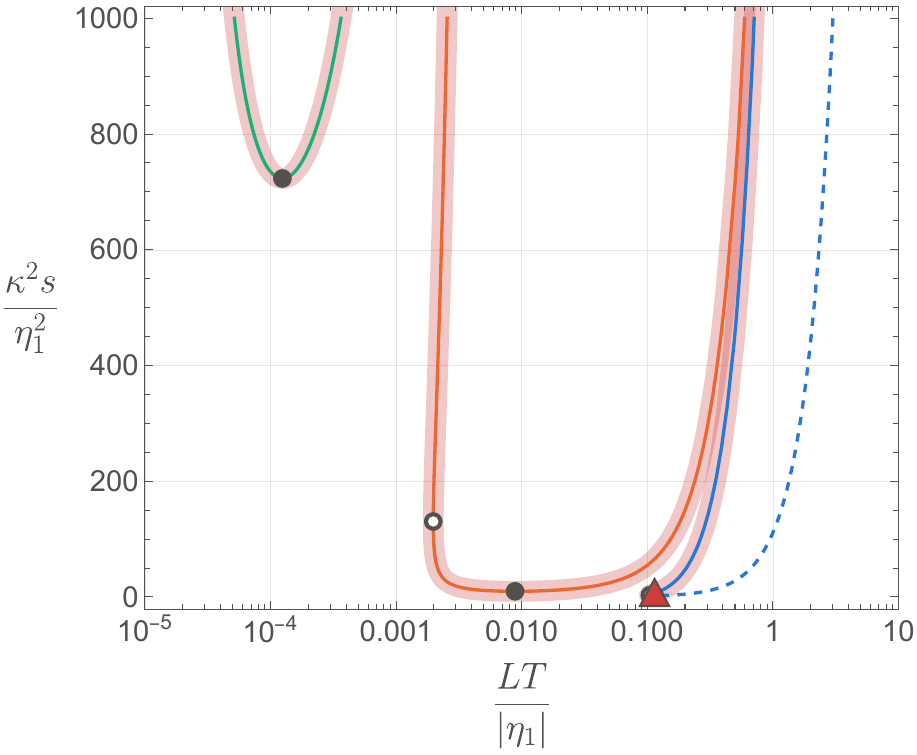}
    \includegraphics[width=0.49\linewidth]{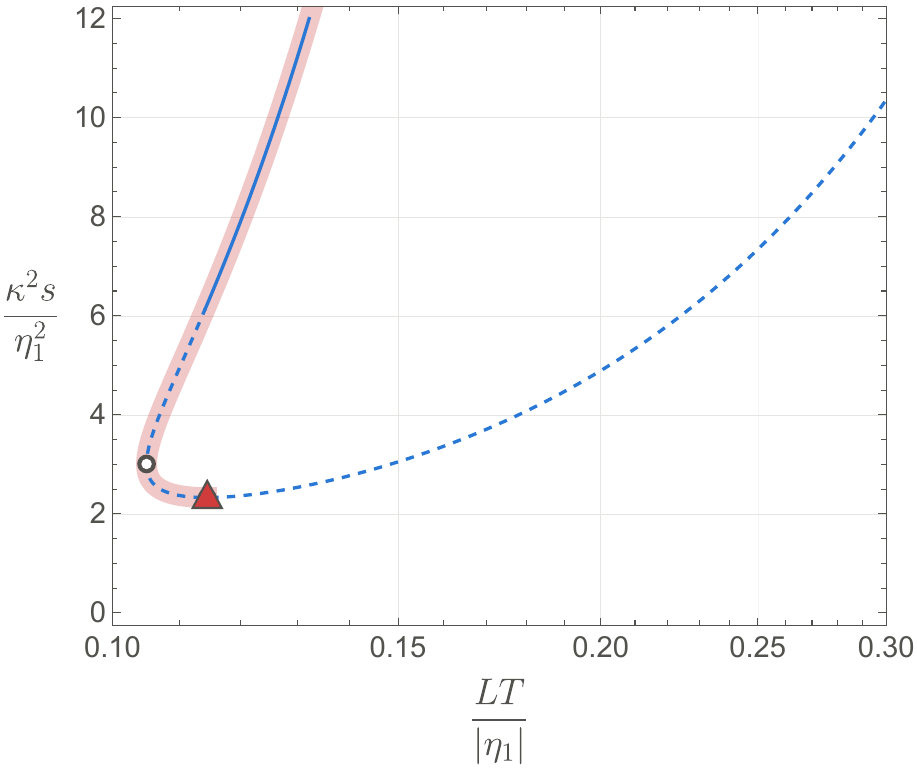}
    
    \caption{Plots of the free energy (top row) and entropy density (bottom row) as a function of the temperature, in units of the source, for the supergravity solutions. The blue, red and green lines correspond to branches with zero, one and two bounces respectively. A black dot marks the point where the specific heat becomes zero and a white dot where it goes to infinity. The specific heat is negative in between. The dashed line is the branch where the throat is behind the horizon, and solid lines are branches where the throat is outside the horizon. Branches with unstable modes are shaded in red, a red triangle marks the point where a purely imaginary mode crosses the real axis. The right plots zoom on the branch without bounces, note that the red triangle is on top of the black dot, an unstable mode appears where the specific heat changes sign.}
    \label{fig:thermoSUGRA}
\end{figure}

\begin{figure}
    \centering
    \includegraphics[width=0.55\linewidth]{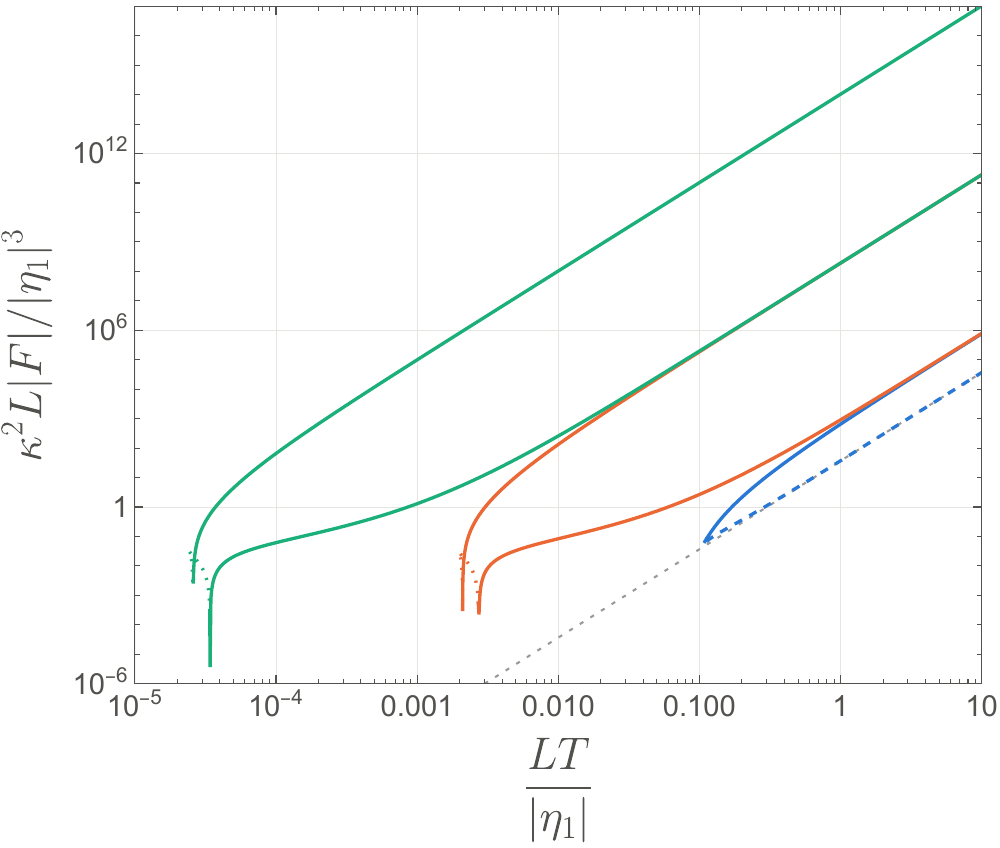}
\caption{Comparison of the absolute value of the free energies of different branches as a function of temperature. The blue, red and green lines correspond to the solutions with no bounces, one bounce and two bounces, respectively. For the red and green lines, the solid lines correspond to negative values and the dashed lines to positive values. For the blue line, the free energy is always negative and solid and dashed distinguishes solutions with the throat outside or behind the horizon. The gray dashed line is the free energy of the AdS black brane solution.}
\label{fig:freeEnComp}
\end{figure}

We have plotted the free energy and entropy densities in Figures~\ref{fig:thermoSUGRA} and \ref{fig:freeEnComp}.
One finds that for each value of $L T/\eta_1$ there are several solutions with different free energies. These solutions split into branches that can be distinguished by the number of bounces. A new bounce appears each time the source $\eta_1$ crosses zero, so consecutive branches connect at $LT/|\eta_1|\to \infty$.
Solutions in the branch with no bounces and the throat behind the horizon at temperatures higher than $LT/|\eta_1|\approx 0.12834$ do not have dS behind the horizon. Since this behavior does not affect the thermodynamics, in the following we will only refer to the bounces between the AdS boundary and the horizon.

Each of the branches with a fixed number of bounces consists of two sub-branches that join at a turning point, located at a minimal value of the temperature that is smaller as the number of bounces increases. One of the sub-branches has higher free energy than the other, and this one has a region with negative specific heat close to the turning point. Comparing branches with different numbers of bounces, the free energy becomes lower as the number of bounces is increased, possibly related to the ghost kinetic term contributing with a negative sign to the Euclidean action in the gravitational theory. This would make the branches with more bounces thermodynamically favored, according to the usual criterion, and any branch with a finite number of bounces and positive specific heat metastable, before linear stability is considered.

\section{Linear stability}\label{sec:qnms}

In order to study the dynamical stability, we expand the equations of motion \eqref{eq:eoms} about the background solutions to linear order in the fluctuations of the metric and the scalar
\begin{equation}
    G_{MN}=g_{MN}+h_{MN}
    \ , \qquad
    \Phi=\im(\eta+\chi)\ .
\end{equation}
The background metric $g_{MN}$ has the form given in \eqref{eq:metricansatz}. For the fake supergravity solutions, the background scalar $\Phi=\im\eta(\rho)$ is given by \eqref{eq:scalarsol}, whereas for the supergravity solutions it is obtained numerically.
Although we have expressions for general dimensions in the fake supergravity models, we restrict the analysis to $d+1=4$ dimensional geometries.

\subsection{Equations of motion}
\label{subsec:EOMs}

We choose to work in the radial gauge and expand the fluctuations in plane waves along the directions parallel to the AdS boundary
\begin{equation}
    h_{M\rho}=0
    \ ,\qquad 
    X(\rho,t,\bm{x})=\int \frac{\dd\omega\, \dd^2 k}{(2\pi)^3}\, X(\rho,\omega,\bm{k})\,e^{-\im\omega t+\im\bm{k}\cdot \bm{x}}\ , 
\end{equation}
where $X=h_{\mu\nu},\ \chi$. After gauge fixing, there is still a residual diffeomorphism transformation.
\footnote{
Specifically, the residual diffeomorphism is 
\begin{align}
    \xi_\rho(\rho,x) 
    =
    \sqrt{f(\rho)}\, \lambda(x) 
    \ , \qquad
    \xi_0(\rho,x)
    = 
    b_0(\rho)\, \partial_0 \lambda(x) 
    \ , \qquad
    \xi_i(\rho,x)
    = 
    b(\rho)\, \partial_i \lambda(x)\ , 
\end{align}
where $\lambda(x)$ is the gauge parameter function and $x=\left\{x^0,x^1,x^2\right\}$ entails only the coordinates along the boundary. The functions $b_0(\rho)$ and $b(\rho)$ are required to satisfy 
\begin{align}
    b_0'(\rho) 
    = 
    \frac{1}{e^{2A(\rho)}f(\rho)^{3/2}}
    \ , \qquad
    b'(\rho) 
    = 
    \frac{1}{e^{2A(\rho)}f(\rho)^{1/2}}\ ,
\end{align}
in order to preserve the radial gauge fixing.}
Without loss of generality, we can take the spatial momentum to be aligned along the $x^1$ direction. Then, the following combinations of fluctuations are invariant under the residual diffeomorphisms:
\begin{equation}\label{eq:fluctuations}
\begin{array}{l|cl}
\text{vector}  & \ & h_{t2}\, ,\ h_{12}\\
\text{metric scalar}\ & \ & Z_2=\omega^2\, h_{11}+k^2\, h_{tt}+2\omega k\, h_{t1}-\left( \omega^2-k^2  \frac{f'}{2A'}e^{-2A}\right)h_{22} \\
\text{scalar} & \ & Z_\phi=\chi-\frac{\eta'}{2A'}e^{-2A} h_{22}
\end{array}
\end{equation}
We look for solutions that are normalizable at the boundary, ingoing at the horizon, and regular in between. These only exist for a discrete set of complex values of the frequency. If the imaginary part of the frequency is non-positive, the fluctuations decay in time or oscillate and the background geometry is dynamically linearly stable. Otherwise, the fluctuations grow exponentially in time and the background is unstable. For simplicity, we limit the analysis to modes with zero spatial momentum $k=0$. This misses potential instabilities towards inhomogeneous solutions, but typically we do not expect those to be present in the absence of a charge density and for backgrounds that are dual to theories with positive specific heat. 

The equations of motion for the invariant modes are:
\begin{itemize}
    \item Vector modes:
    \begin{subequations}
    \begin{eqnarray}
        0&=& h_{t2}'-2A' h_{t2}\ ,\\
        0 &=& h_{t2}''-A' h_{t2}' -\left(2 (A')^2+\frac{(\eta ')^2}{2}\right)h_{t2}\ ,\\
        0&=& h_{12}''+ \left(\frac{f'}{f}-A'\right)h_{12}'- \left(2 (A')^2+\frac{(\eta ')^2}{2}+\frac{2 A' f'}{f}-\frac{e^{-2 A}\omega^2 }{f^2}\right)h_{12}\ .
    \end{eqnarray}
\end{subequations}
    \item Metric scalar:
    \begin{equation}\label{eq:Z2}
        0= Z_2''+ \left(\frac{f'}{f}-A'\right)Z_2'- \left(2 (A')^2+\frac{(\eta ')^2}{2}+\frac{2 A' f'}{f}-\frac{e^{-2 A} \omega^2 }{f^2}\right)Z_2\,.
    \end{equation}
    \item Scalar:
    \begin{equation}\label{eq:Zphi}
        0=Z_\phi''+a_{\phi\phi} Z_\phi'+b_{\phi\phi} Z_\phi+\frac{e^{-2 A}}{\omega^2}\left(a_{\phi 2} Z_2'+ b_{\phi 2}Z_2\right)\,.
    \end{equation}
    The coefficients are
    \begin{subequations}
     \begin{eqnarray}
     a_{\phi\phi} &=& 3 A'+\frac{f'}{f}\,,\\
     b_{\phi\phi} &=&-\frac{\left(\eta '\right)^4}{8 \left(A'\right)^2} -\left(\eta '\right)^2 \left(\frac{f'}{2 f A'}+\frac{3}{2}\right)-\frac{\eta ' \partial_\eta V(\eta )}{f A'}+\frac{\partial_\eta^2 V(\eta )}{f}+\frac{e^{-2 A}
   \omega^2}{f^2}\,, \label{eq:bphiphi}\\
   a_{\phi 2} &=&\frac{ \left(\eta '\right)^3}{8 \left(A'\right)^2}+ \eta ' \left(\frac{ f'}{2 f A'}+\frac{3 }{2}\right)+\frac{ V'(\eta )}{2 f A'}\,,\\
   b_{\phi 2} &=& -\left(\eta '\right)^3 \left(\frac{ f'}{16 f \left(A'\right)^2}+\frac{1}{4 A'}\right)-\eta ' \left(\frac{ \left(f'\right)^2}{4 f^2 A'} +3 
   A'+\frac{7  f'}{4 f}\right)\nonumber \\ & &
   -\frac{ f' V'(\eta )}{4 f^2 A'}-\frac{ V'(\eta )}{f}\,.
     \end{eqnarray}
    \end{subequations}
\end{itemize}
The mode $h_{t2}$ is not dynamical, and can be set to zero. This gives three possible independent dynamical fluctuations. The modes $h_{12}$ and $Z_2$ satisfy the same decoupled equations, so it is enough to study one of them. The mode $Z_2$ sources $Z_\phi$, but solving the equation for $Z_2$ is enough to determine stability. On the other hand, since $Z_\phi$ does not source $Z_2$, it is consistent to set $Z_2=0$ in \eqref{eq:Zphi} to complete the study of the stability of all the possible dynamical modes.

There is an important difference between the equation of the metric fluctuations \eqref{eq:Z2} and the equation of the scalar fluctuation \eqref{eq:Zphi}. Both have a regular singular point at the horizon, where $f$ has a simple zero. However, the equation of the scalar \eqref{eq:Zphi} has an additional regular singular point at the throat, from the first term in the coefficient $b_{\phi\phi}$ in \eqref{eq:bphiphi}, unless the minimum is exactly at the horizon. At this point the fluctuation $Z_\phi$ can be singular, due to the $1/A'$ factor appearing in \eqref{eq:fluctuations}. However, the scalar and metric fluctuations remain regular.

\subsection{Spectrum of fluctuations in fake supergravity models}

The calculation of the spectrum of fluctuations depends on the relative position of the throat and the black hole horizon, as one has to deal with the additional singular point on the equation of $Z_\phi$. More specifically, referring to \eqref{eq:eA}, \eqref{eq:f2} and \eqref{eq:rhoh}, the throat falls outside the horizon for $\frac{1}{2}<f_0<1$, being instead cloaked by it for $f_0>1$. For the special case $f_0=1$, the radial position of the throat and the horizon coincide. 

\begin{figure}
    \centering
    \includegraphics[width=0.49\linewidth]{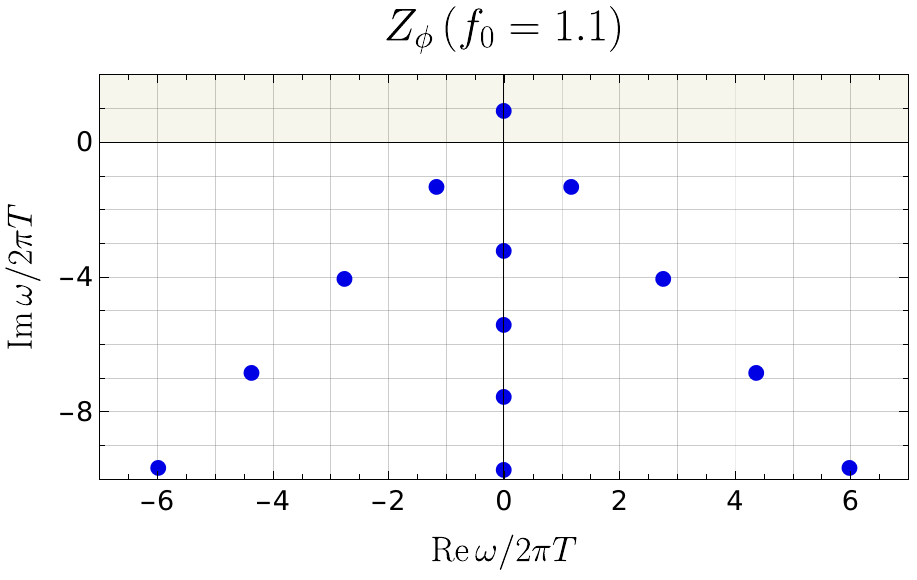}
    \includegraphics[width=0.49\linewidth]{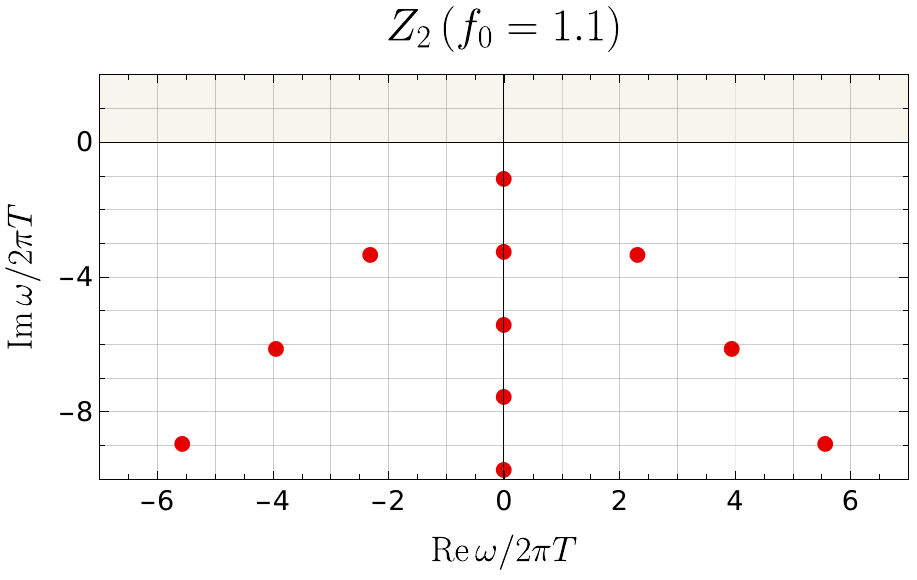}
    \includegraphics[width=0.49\linewidth]{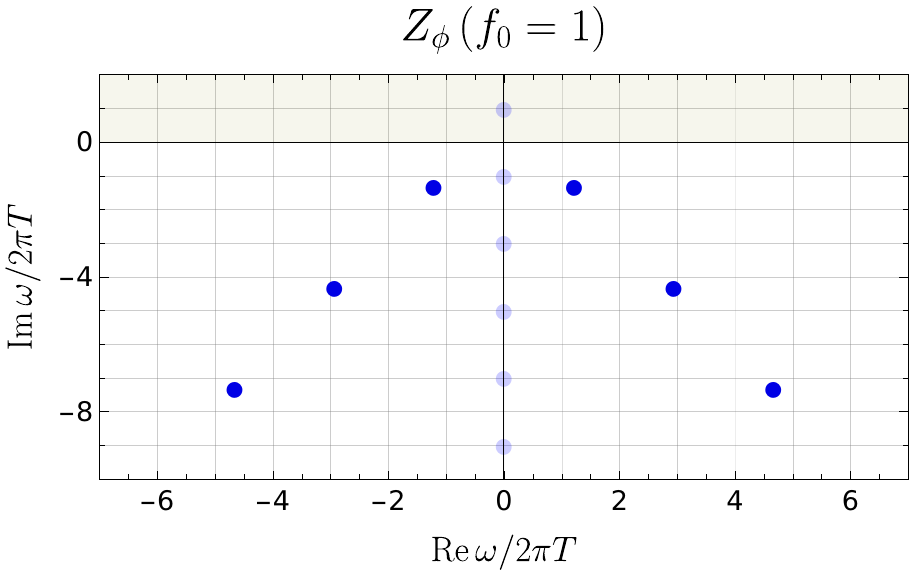}
    \includegraphics[width=0.49\linewidth]{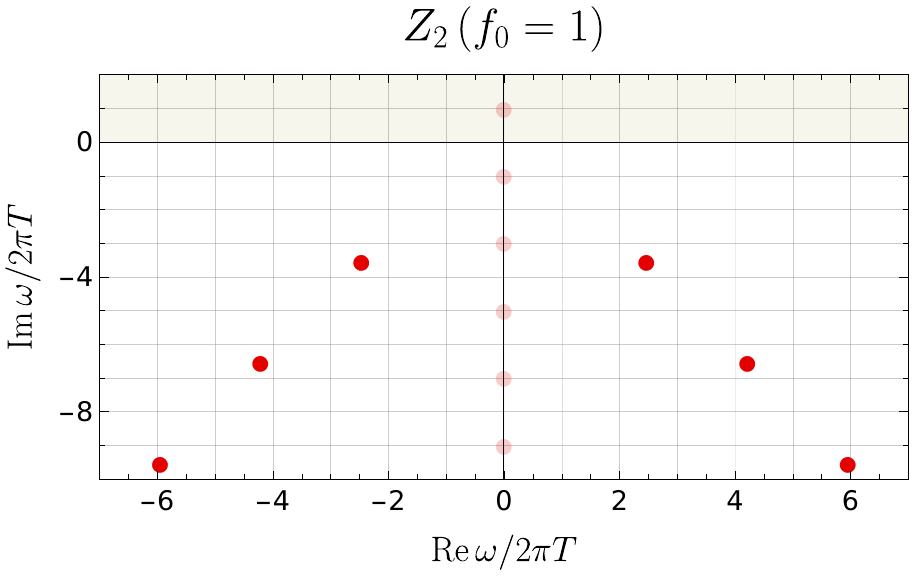}
    \includegraphics[width=0.49\linewidth]{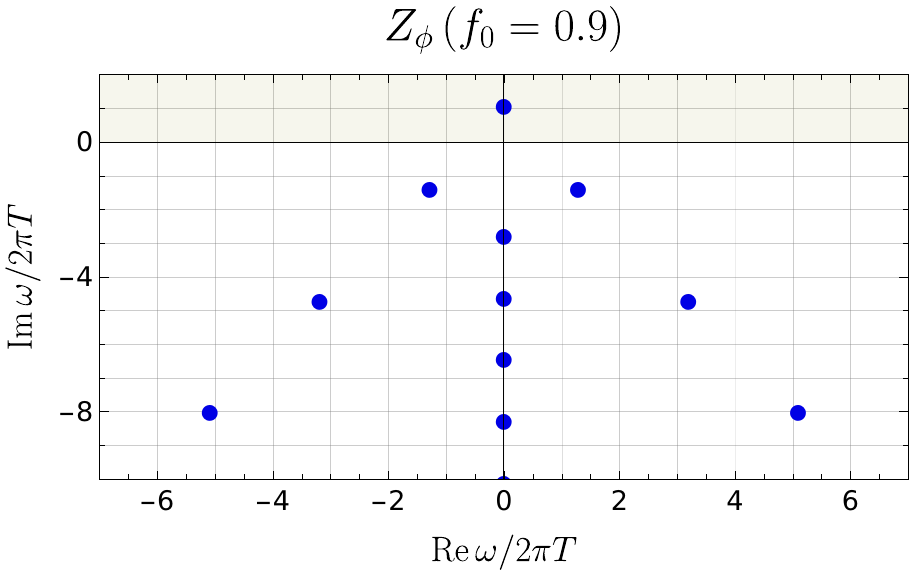}
    \includegraphics[width=0.49\linewidth]{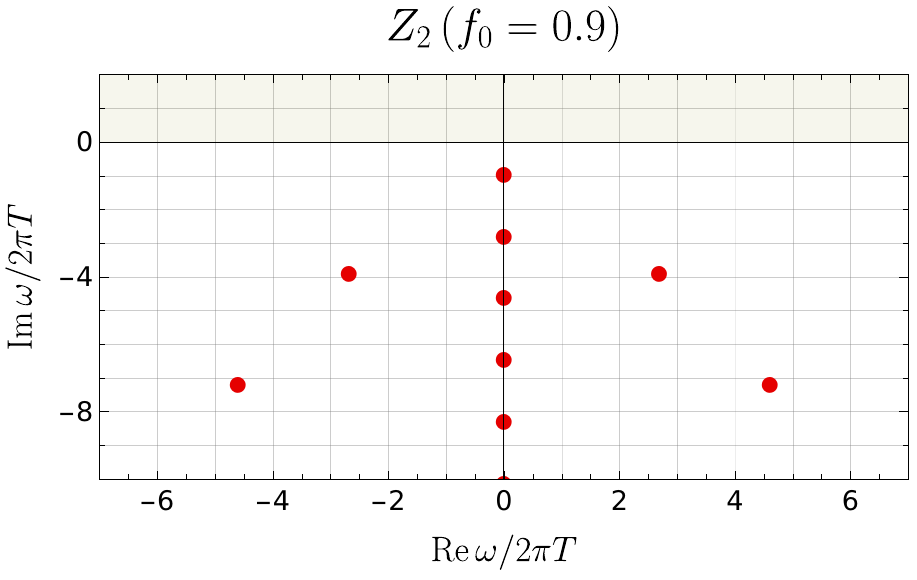}
    
    \caption{Lowest QNMs for the scalar combinations $Z_\phi$ (left) and $Z_2$ (right) depending on the relative position of the throat and the horizon. Frequencies are normalized as $\mathfrak{w}=\frac{\omega}{2\pi T}$. Top: $f_0 > 1$, the throat is behind the horizon; $Z_\phi$ features an unstable mode. Center: $f_0 = 1$, the throat sits on the horizon. The paler dots on the imaginary axis are located at $\mathfrak{w}_n=-(2n-1)\, \im$ with $n=0,1,2,3,...$ and correspond to pole skipping (for $n=0$ the dot lies in the upper half-plane, $\mathfrak{w}_0=+\im$). Bottom: $f_0 < 1$, the throat is outside the horizon. $Z_\phi$ presents again an unstable mode. $Z_2$ has a stable spectrum for all values of $f_0$.}
    \label{fig:QNM}
\end{figure}

When the throat is inside or at the horizon, $f_0\geq 1$, the calculation of the quasinormal mode (QNM) spectrum of $Z_2$ and $Z_\phi$ proceeds in a similar way. One imposes that the fluctuations are normalizable at the boundary and ingoing at the horizon, thus finding the values of the frequency for which this is possible. The details of the calculation can be found in Appendix~\ref{app:qnms} while the results are depicted in Figure~\ref{fig:QNM}.

When the throat is outside the horizon, $\frac{1}{2}< f_0<1$, a new singular point appears at the radial location of the throat in the equation of motion for $Z_\phi$, as mentioned before. This issue can be avoided by defining a new gauge-invariant fluctuation $W_\phi=A'Z_\phi$, which is regular at the throat and leads to a well-defined quasinormal spectrum; the details of the calculation can be found in Appendix~\ref{app:qnms}.
The spectrum is shown in Figure~\ref{fig:QNM}. The analysis of the QNMs of $Z_2$, instead is analogous independently of the relative position of the throat and the horizon.

\begin{figure}[t!]
    \centering
    \includegraphics[width=0.6\linewidth]{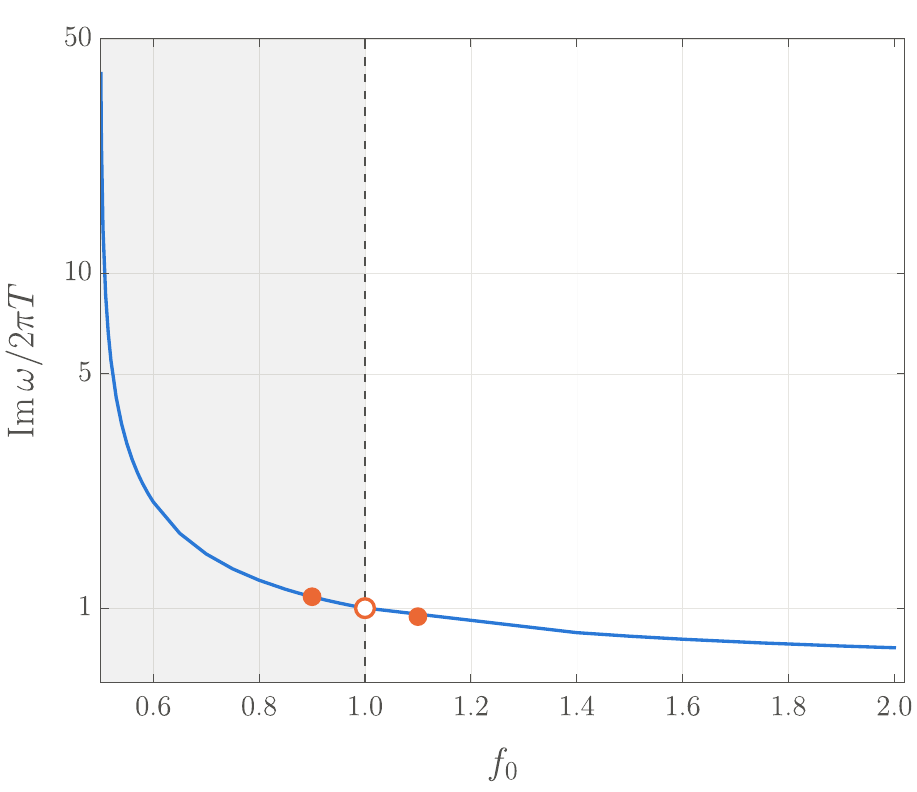}

\caption{Imaginary part of the unstable mode in the fake supergravity models as a function of $f_0$. In the gray area the throat is cloaked by the horizon and in the white area it is outside. The dots mark the values for which the spectrum in Figure~\ref{fig:QNM} is computed. The empty dot is where the throat and the horizon coincide and the unstable pole is skipped.}\label{fig:fakeins}

\end{figure}

Linear stability can be inferred from the quasinormal spectra by looking at the imaginary part of the QNM frequencies. For frequencies that are real or have a negative imaginary part, the modes are purely oscillatory or decay exponentially in time, thus the backgrounds are linearly stable with respect to these fluctuations. On the other hand, a positive imaginary part implies that the mode grows exponentially and the background is linearly unstable. The results for the spectra show that there is an unstable mode in the scalar fluctuation $Z_\phi$ for all values of $f_0$ that we have studied. More precisely, at $f_0=1$, the pole skipping phenomenon occurs \cite{Blake:2018leo,Grozdanov:2019uhi}, but there is still an unstable mode that is regular at the future horizon. It corresponds to the pale dot in the unstable half-plane for $Z_\phi$ in Figure~\ref{fig:QNM}.
In Figure~\ref{fig:fakeins} we track the unstable mode for different values of $f_0$.

\subsection{Spectrum of fluctuations in the supergravity model}

The calculation of the spectrum is qualitatively similar to the one we discussed for the fake supergravity models. The stability for the cases we have examined, up to two bounces, is summarized in Figure~\ref{fig:thermoSUGRA}. Let us describe the behavior as one moves from right to left along each of the curves of the free energy. 

One starts on the branch with no bounces and the throat behind the horizon. It turns out that this branch is linearly stable at high temperatures, until a point where the specific heat vanishes (black dot) and one enters a thermodynamically unstable region. At that point, a mode crosses the real frequency axis from negative to positive imaginary frequencies (red triangle). The instability persists even when the thermodynamically unstable region is exited at the turning point (white dot). The instability remains when the throat is at the horizon. From that point on, the solutions with the throat outside the horizon are also unstable. 

The solutions with the throat outside the horizon connect at $LT/|\eta_1|\to \infty$ with the branch with one bounce, which inherits the instability and remains unstable for the whole branch, independently of the sign of the specific heat. A calculation of the branch with two bounces also reveals that it is unstable everywhere. It seems likely that branches with a larger number of bounces will also be unstable.

\section{Discussion}\label{sec:discuss}

We have constructed explicit solutions interpolating between asymptotic AdS and dS regions in more than two dimensions. The AdS and dS asymptotic regions are separated by a horizon. Note that, in two dimensions, one can identify a patch of dS outside the horizon. Here such patch might correspond to a region outside the horizon where the scalar potential is positive, although the presence of scalar gradients makes the identification less sharp than in two dimensions. Having an identifiable extended dS region outside the horizon may require further adjustments of the potential.

The solutions admit a holographic interpretation as thermal states of a pseudo-Hermitian theory, obtained by the continuation of a real scalar operator and associated coupling in the dual field theory to imaginary values. In one set of examples the solutions are constructed using a fake supergravity approach and have no known embedding in string theory. The other example is based on a consistent truncation that can be uplifted to eleven-dimensional supergravity and therefore belongs to the low energy description of M-theory. 

We have studied the thermodynamic and linear stability of the solutions in the fake supergravity models, and found AdS to dS solutions, with no sources in the dual field theory, and a free energy that is always lower than that of an empty AdS black brane. They have positive specific heat so they are thermodynamically stable, but they are not linearly stable.
 
While the geometry appears smooth in the fake supergravity models and in the consistent truncation, the situation is different when one studies the embedding of the latter in string theory. The uplift of the supergravity solution to eleven dimensions reveals some issues that are invisible in the consistent truncation; at the location of the throat  some fluxes and components of the metric diverge, and a jump in the derivative of some metric components signals the presence of a thin brane. This seems to be consistent with the imaginary distance conjecture \cite{Maldacena:2026jqd, DiUbaldo:2026rly, Etheredge:2026rio} (see \cite{Begines:2026mlv} for massive fields) predicting a breakdown of the low-energy effective theory when the scalar field extended to imaginary values reaches the value at the throat. In this case the modes captured by the consistent truncation are not sensitive to the singular behavior at the throat, and the issues predicted by the conjecture are only observed when other sectors of the low-energy effective theory are considered. It would be interesting to understand better why this is the case, and whether the singularity at the throat is admissible in the full theory, in the sense of having a consistent interpretation in terms of M-theory sources. We leave this analysis for future work.

Regarding the thermodynamical and linear stability of the supergravity model within the consistent truncation, the result is unclear. We find AdS to dS solutions that have positive specific heat and are linearly stable. However, there seems to be an infinite family of unstable solutions with lower free energies, for any given value of the temperature in units of the source where the stable solutions exist. If all the unstable solutions were discarded, the stable solutions would be those with the highest free energy. One might entertain the possibility that, in a system with a Hamiltonian that is bounded {\em from above}, the proper thermal density matrix operator should be $\hat\varrho=\frac{1}{Z}e^{\hat{H}/T}$, with $Z={\rm Tr}(e^{\hat{H}/T})$ the partition function, akin to having a negative temperature $T_{\rm eff}=-T$. Then, at the saddle points of the thermal path integral, the logarithm of the partition function would be the classical action, rather than minus the classical action. Effectively this would reverse the sign of the free energy, so that the more unstable solutions would have larger free energy and the no-bounce solutions with the throat behind the horizon would be the thermodynamically preferred state. If this prescription were applied to the fake supergravity solutions, it would make the empty AdS black brane the thermodynamically preferred state. In both instances the thermodynamically preferred state would also be the linearly stable state. We leave as an open question if this is an admissible interpretation of the solutions from the point of view of the holographic dual. Note also that probably there are missing configurations, since the stable part of the no-bounce branch does not cover the full temperature range. It would be interesting to find additional solutions that complete the phase diagram.

Finally, although we have not studied explicitly fluctuations inside the horizon, it should be noted that the linearized ghost scalar equation is similar to that of an ordinary scalar of positive mass. The solutions then decay at the asymptotic dS boundary, similarly to how they decay at the asymptotic AdS boundary. In addition, the solutions for the scalar and the metric fluctuations are regular at the throat. We thus do not expect any linear instabilities in this region. 

In conclusion, we have shown that linearly stable AdS to dS solutions exist in holographic duals to pseudo-Hermitian theories obtained through the continuation to imaginary values of real couplings and operators. Their global stability is however unclear, since a proper thermodynamic formulation may be different from the one applied to the original Hermitian theories. Further issues in the low energy description emerge when considering an embedding in string theory. Addressing these questions would be a necessary step towards a consistent holographic formulation of these theories.

\section*{Acknowledgements}

We thank Daniel Are\'an, Suman Das, David García-Fariña and Juan Pedraza for useful comments and discussions. This work is partially supported by the Spanish Agencia Estatal de Investigación and Ministerio de Ciencia, Innovaci\'on y Universidades through the grant PID2024-161500NB-I00. We used Claude Opus 5.5 to check the fluctuation equations and the QNM spectra in the fake supergravity models; to produce and check some of the numerical results in the supergravity models; and to proofread the draft.

\appendix

\section{Numerical backgrounds of the supergravity model}\label{app:numericsSUGRA}

In this appendix we will summarize the calculations performed to obtain the numerical backgrounds in Section~\ref{sec:sugra}. These are solutions to equations \eqref{eq:eomsansatz} 
satisfying the asymptotic behaviors described in \eqref{eq:asympcond}.
The expansion of the solutions around $\rho\rightarrow-\infty$  to leading order is given in \eqref{eq:rhoexpansions}, and parametrized by 6 constants ($f_{-\infty},f_3,\eta_1,\eta_2,a_{-\infty},a_0$) that appear as the coefficients of the leading terms in the expansion:
\begin{align}
    f&=f_{-\infty}+f_3 \,e^{\frac{3a_0\rho}{L}}+O\left(e^{\frac{5a_0\rho}{L}}\right)\,,\nonumber\\[2mm]
    \eta&=\eta_1\, e^{\frac{a_0\rho}{L}}+\eta_2\, e^{\frac{2a_0\rho}{L}}+O\left(e^{\frac{3a_0\rho}{L}}\right)\,,\\[2mm]
    A&=a_{-\infty}-a_0\frac{\rho}{L}+a_1e^{\frac{a_0\rho}{L}}+a_2e^{\frac{2a_0\rho}{L}}+O\left(e^{\frac{3a_0\rho}{L}}\right)\,,\nonumber
\end{align}
where the subleading terms, including $a_1$ and $a_2$, will be fully determined by the integration constants. Using a change of coordinates one can set $a_0=1$, as we will explain below. 

If we insert this expansion in the equations of motions, taking into account the potential \eqref{eq:VsugraIm}, we obtain the conditions:
\begin{align}
  0 &= 4a_0^2\,a_1\,e^{\frac{a_0\rho}{L}} + a_0^2\left(16a_2-\eta_1^2\right)e^{\frac{2a_0\rho}{L}} + O\!\left(e^{\frac{3a_0\rho}{L}}\right),\\[2mm]
  0 &= O\!\left(e^{\frac{4a_0\rho}{L}}\right),\\[2mm]
  0 &= 6\left(a_0^2f_{-\infty}-1\right) - 6a_0^2a_1f_{-\infty}\,e^{\frac{a_0\rho}{L}}
            + \Big[6a_0^2a_1^2f_{-\infty} + \eta_1^2\left(1-a_0^2f_{-\infty}\right)\Big]e^{\frac{2a_0\rho}{L}} + O\!\left(e^{\frac{3a_0\rho}{L}}\right),\\[2mm]
  0 &= 2\eta_1\left(\frac{1}{f_{-\infty}}-a_0^2\right)e^{\frac{a_0\rho}{L}}
            + \left[3a_0^2a_1\eta_1 + 2\eta_2\left(\frac{1}{f_{-\infty}}-a_0^2\right)\right]e^{\frac{2a_0\rho}{L}} + O\!\left(e^{\frac{3a_0\rho}{L}}\right).
\end{align}
The equations leave $f_3,\eta_1,\eta_2$ and $a_{-\infty}$ undetermined. The first equation 
fixes the coefficients of the subleading terms in $A$ in terms of the coefficients of the leading terms in $\eta$:
\begin{equation}
    a_1=0\,,\quad
    a_2=\frac{\eta_1^2}{16}\,.
\end{equation}
The fourth equation relates two of the coefficients of the leading terms:
\begin{equation}
    f_{-\infty}=\frac{1}{a_0^2}\ .
\end{equation}
The third equation is a constraint, so once we have fixed the subleading coefficients using the other equations, it vanishes order by order.

If we consider the same but for the asymptotic behavior at $\rho\rightarrow+\infty$, we obtain the same results, because it is really the same problem under a symmetry transformation:
\begin{equation}
    \eta\rightarrow\sqrt{3}\pi-\eta ~~,~~ \rho\rightarrow-\rho ~~,~~ f\rightarrow-f\ .
\end{equation}

The desired solution would have $a_0=1$. However, to ensure continuity, we will ``shoot'' the numerical solution from the horizon, so we cannot directly fix the boundary conditions at $\rho\to-\infty$. What we can do a posteriori to have the desired normalization is the following coordinate transformation:
\begin{equation}\label{eq:normalization}
        \rho = \frac{\bar{\rho}}{a_0} ~~,~~ t= e^{-a_{-\infty}/2}a_0\bar{t} ~~,~~ \bar{f}(\bar{\rho}) = \frac{1}{g_{\bar{\rho}\bar{\rho}}} = a_0^2 f\left(\frac{\bar{\rho}}{a_0}\right) ~~,~~ \mathbf{x} = e^{-a_{-\infty}/2}\mathbf{\bar{x}}\ .
\end{equation}
There is one additional caveat: we cannot calculate global solutions because $f(\rho)$ will acquire a null value at the horizon which constitutes a singularity in the equations. To get around this, as we previously mentioned, we will consider boundary conditions asymptotically close to the horizon from both sides, so that the full solution and its derivatives are continuous.

We choose the $\rho$ coordinate in such a way that the horizon is at $\rho=0$. 
Then, we expand assuming that the solution is regular at the horizon:
\begin{align}
  f &= F_1\rho + F_2\rho^2 + F_3\rho^3 + F_4\rho^4 + O(\rho^5)\,,\nonumber\\
  A &= A_H + A_1\rho + A_2\rho^2 + A_3\rho^3 + A_4\rho^4 + O(\rho^5)\,,\\
  \eta &= \eta_H + H_1\rho + H_2\rho^2 + H_3\rho^3 + O(\rho^4)\,.\nonumber
\end{align}
If we insert this expansion in the equations of motion we get the following conditions up to second order:
\begin{align}
    0 &= \left(8A_2-H_1^2\right) + 4\left(6A_3-H_1H_2\right)\rho + \left(48A_4-4H_2^2-6H_1H_3\right)\rho^2,\\[2mm]
    0 &= \left(2F_2+3A_1F_1\right) + 6\left(F_3+A_2F_1+A_1F_2\right)\rho + 3\left(4F_4+3A_3F_1+4A_2F_2+3A_1F_3\right)\rho^2,\\[2mm]
    0 &= \left(2A_1F_1-6c\right) + \left[6A_1^2F_1+4A_1F_2+F_1\big(4A_2+\frac{H_1^2}{2}\big)+2\sqrt3\,s\,H_1\right]\rho \nonumber\\
    &\quad + \Big[6A_3F_1+6A_1^2F_2+8A_2F_2+6A_1\big(4A_2F_1+F_3\big)+\frac{F_2}{2}H_1^2 \\
    &\quad +2F_1H_1H_2 + c\,H_1^2 + 2\sqrt3\,s\,H_2\Big]\rho^2,\nonumber\\[2mm]
    0 &= \left(F_1H_1+2\sqrt3\,s\right)
     + \Big[3A_1F_1H_1+2F_2H_1+4F_1H_2+2c\,H_1\Big]\rho \nonumber\\
    &\quad + \Big[3\big(2A_2F_1H_1+A_1F_2H_1+F_3H_1+2A_1F_1H_2+2F_2H_2+3F_1H_3\big) \\
    &\quad+2c\,H_2-\tfrac{s}{\sqrt3}\,H_1^2\Big]\rho^2 \nonumber\,,
\end{align}
where we have introduced:
\begin{equation}
    c = \frac{1}{L^2}\cos\left(\frac{\eta_H}{\sqrt{3}}\right) ~~,~~ s = \frac{1}{L^2}\sin\left(\frac{\eta_H}{\sqrt{3}}\right) \,.
\end{equation}
From the above conditions, we can obtain the coefficients of the solution solving order by order:
\begin{align}
    \mathcal{O}(1):&\quad A_1=\frac{3c}{F_1},\quad A_2=\frac{3s^2}{2F_1^2},\quad F_2=-\frac{9c}{2},\quad H_1=-\frac{2\sqrt3\,s}{F_1}\,,\\[2mm]
    \mathcal{O}(\rho):&\quad A_3=-\frac{c\,s^2}{F_1^3},\quad F_3=\frac{3\left(4+5\cos\frac{2\eta_H}{\sqrt3}\right)}{2F_1L^4},\quad H_2=\frac{\sqrt3\,\sin\frac{2\eta_H}{\sqrt3}}{2F_1^2L^4}\,,\\[2mm]
    \mathcal{O}(\rho^2):&\quad A_4=
    \frac{\left(9\cos\frac{2\eta_H}{\sqrt3}-17\right)s^2}{24F_1^4L^4},\quad
    F_4=-\frac{3\left(107L^2\,c+55\cos(\sqrt3\,\eta_H)\right)}{16F_1^2L^6},\quad\\
    &\quad H_3=\frac{\left(10-3\cos\frac{2\eta_H}{\sqrt3}\right)s}{3\sqrt3\,F_1^3L^4}\nonumber\,.
\end{align}
One thing worth noticing is that $F_1:=f'(0)$ only appears as a scale factor, so its value will only determine the scale of $\rho$ with respect to the structure of the solution. This is a consequence of the symmetry manifest in the change of coordinates in equation \eqref{eq:normalization}: rescaling $f$ (and consequently $f'$) by a constant can be absorbed by a rescaling of the radial coordinate. Also, $A_H$ does not influence the solution at all, its value constitutes just a vertical shift of the $A$ profile. All of this tells us that the numerical solution is determined by a single parameter, which will be the value of the scalar at the horizon $\eta_H:=\eta(0)$.

We then solve the equations numerically using the horizon expansion to give initial conditions, with a fixed choice of $A_H$ and $F_1$, which do not actually influence the structure of the solution; and a varying choice of $\eta_H$. After we have obtained the numerical solutions, we just normalize them appropriately following the change of coordinates in equation \eqref{eq:normalization}. We plot some examples in Figure~\ref{fig:solSUGRA}.

\section{Holographic renormalization}\label{app:holoren}

 The free energy and other thermodynamic quantities can be computed from the on-shell action. In order to regularize the result, we need to introduce the York--Gibbons--Hawking term and an appropriate counterterm. The total on-shell Euclidean action is thus
\begin{align}
S_{\text{os}}
&=-\frac{1}{2\kappa^2}\int\dd^{d+1}x\, \sqrt{G}\left[R-\frac12\partial_M\Phi\partial^M\Phi-V(\Phi)\right]
\\ \nonumber &\qquad 
-\frac{1}{\kappa^2}\int\dd^dx\, \sqrt{\gamma}\,K+\frac{1}{2\kappa^2}\int\dd^dx\, \sqrt{\gamma}\,W(\Phi)\,,
\end{align}
where $\gamma$ is the induced metric on the boundary and $K$ is the trace of the extrinsic curvature. If a superpotential exists, the counterterm $W$ can be made equal to it, assuming the solution it determines has the leading boundary fall-off.

By means of standard manipulations, the bulk part of the action can be written as a total derivative of the temporal component of the extrinsic curvature. After performing the volume integral on the boundary one gets\footnote{Note that, being the boundary at $\rho\to-\infty$, the radial lower and upper integration limits are $\rho_{\text{\tiny UV}}$ and $\rho_{h}$ respectively.} 
\begin{align}
S_{\text{os}}=\left.\frac{\beta V_{d-1}}{\kappa^2}\left(\sqrt{\gamma}K^t{}_t\right)\right|_{\rho_{h}}-\frac{\beta V_{d-1}}{\kappa^2}\lim_{\rho_{\text{\tiny UV}}\to-\infty}\left[\sqrt{\gamma}\left(K^t{}_t-K+\frac12W\right)\right]_{\rho_{\text{\tiny UV}}}\,,
\end{align}
where $\rho_{\text{\tiny UV}}$ is some cutoff needed to regularize the action, $V_{d-1}$ is the (infinite) spatial volume and $\beta$ is the period of the Euclidean time coordinate, that is, the inverse of the temperature. The first term is evaluated at the horizon, so it is absent when there is none. Using this, the free energy density is finally
\begin{align}\label{eq:free}
F&=\frac{S_{\text{os}}}{\beta V_{d-1}}=\left.\frac{1}{\kappa^2}\left(\sqrt{\gamma}K^t{}_t\right)\right|_{\rho_h}-\frac{1}{\kappa^2}\lim_{\rho_{\text{\tiny UV}}\to-\infty}\left[\sqrt{\gamma}\left(K^t{}_t-K+\frac12W\right)\right]_{\rho_{\text{\tiny UV}}}\nonumber\\[2mm]
&=\frac{1}{\kappa^2}\left[e^{dA}\left(f\,A'+\frac12f'\right)\right]_{\rho_h}-\frac{1}{\kappa^2}\lim_{\rho\to-\infty}\left[\sqrt{f}\,e^{dA}\left(\frac12W-(d-1)\sqrt{f}\,A'\right)\right]\,.
\end{align}
The first term is precisely (minus) the product of the entropy density times the temperature, while the second one should coincide with the internal energy. For pure AdS, with $\Phi=0$, $f=1$ and $A=-\frac{\rho}{L}$, this free energy vanishes for the appropriate choice of $W$.\footnote{This is simply $W=\frac{2(d-1)}{L}$ for a vanishing scalar in the examples we consider.} For the AdS black hole, which in this coordinate system reads
\begin{align}
\Phi=0\,,\qquad\qquad A=-\frac{\rho}{L}\,,\qquad\qquad f=1-e^{\frac{d}{L}(\rho-\rho_h)}\,,
\end{align}
the free energy is
\begin{align}
F_{\text{\tiny AdS}}=-\frac{1}{2\kappa^2L}\,e^{-d\frac{\rho_h}{L}}=-\frac{1}{2\kappa^2L}\left(\frac{4\pi L}{d}\right)^dT^d\,,
\end{align}
where we have used that the temperature is $T=\frac{d}{4\pi L}e^{-\frac{\rho_h}{L}}$.

Varying the renormalized on-shell action with respect to the boundary metric it is also possible to compute the internal energy density and pressure. The boundary energy-momentum tensor is given by 
\begin{align}
T^\mu{}_\nu=-\frac{1}{\kappa^2}\lim_{\rho_{\text{\tiny UV}}\to-\infty}\left[\sqrt{\gamma}\left(K^\mu{}_\nu-\delta^\mu{}_\nu\left(K-\frac12W\right)\right)\right]_{\rho_{\text{\tiny UV}}}=\left(\begin{array}{cc}-\varepsilon & 0\\
0 & p\,\mathbb{1}_{d-1}\end{array}\right)
\end{align}
In this way, the internal energy density is
\begin{align}\label{eq:intenergy}
\varepsilon&=-T^t{}_t=-\frac{1}{\kappa^2}\lim_{\rho_{\text{\tiny UV}}\to-\infty}\left[\sqrt{\gamma}\left(K^t{}_t-K+\frac12W\right)\right]_{\rho_{\text{\tiny UV}}}\nonumber\\
&=\frac{1}{\kappa^2}\lim_{\rho\to-\infty}\left[\sqrt{f}\,e^{dA}\left((d-1)\sqrt{f}\,A'-\frac12W\right)\right]\,,
\end{align}
while the pressure reads 
\begin{align}\label{eq:pressure}
p&=T^x{}_x=\frac{1}{\kappa^2}\lim_{\rho_{\text{\tiny UV}}\to-\infty}\left[\sqrt{\gamma}\left(K^x{}_x-K+\frac12W\right)\right]_{\rho_{\text{\tiny UV}}}\nonumber\\[2mm]
&=\frac{1}{\kappa^2}\lim_{\rho\to-\infty}\left[e^{dA}\left(\frac{1}{2}\sqrt{f}W-\frac12\,f'-(d-1)fA'\right)\right]\,.
\end{align}
The AdS black hole values are $\varepsilon_{\text{\tiny AdS}}=\frac{(d-1)}{2\kappa^2L}\,e^{-d\frac{\rho_h}{L}}$ and $p_\text{\tiny AdS}=\frac{1}{2\kappa^2L}\,e^{-d\frac{\rho_h}{L}}$ respectively. One can see that the energy-momentum tensor of the AdS black hole is traceless, as dictated by conformal invariance. On the other hand, the entropy density, computed from the area of the horizon, is $s_{\text{\tiny AdS}}=\frac{2\pi}{\kappa^2}\,e^{-(d-1)\frac{\rho_h}{L}}$, showing that the thermodynamic relation $\varepsilon+p=sT$ is indeed satisfied. This thermodynamic relation can be obtained alternatively from the conserved charge
\begin{align}\label{eq:conscharge}
   Q(\rho)=- \frac{1}{2\kappa^2}e^{dA(\rho)}f'(\rho)\,.
\end{align}
Using the equation of motion \eqref{eq:eomsansatzf}, it is easy to check that $Q'=0$. At the horizon, it takes the value $Q(\rho_h)=sT $, while evaluated at the AdS boundary it gives $Q(-\infty)=\varepsilon+p$.

\section{Numerical calculation of the quasinormal modes}\label{app:qnms}

We performed the computations of the QNMs with pseudospectral methods. More specifically, we discretized the radial direction on a Gauss--Lobatto collocation grid and adopted a basis of Chebyshev polynomials. The results reported in the main text and Figure~\ref{fig:QNM} have been obtained by means of a grid of 200 collocation points. We nevertheless tested the stability of the results with respect to the number of discretization points. As a further source of checks, we obtained the few lowest modes with an independent shooting method.
\footnote{
An advantage of the shooting method is that it allows to identify which of the purely imaginary modes obtained through the pseudospectral methos are skipped  at $f_0=1$ (see Figure~\ref{fig:QNM}).
}

\subsection{Scalar fluctuations: throat at the horizon}
\label{subsec:atthehor}

This situation corresponds to $f_0 = 1$. The radial variable $\rho$ used in the equations of motion in Subsection~\ref{subsec:EOMs} spans the interval $(-\infty,\rho_h)$ corresponding, respectively, to the boundary and the horizon. According to \eqref{eq:rhoh}, one can set $\rho_h = 0$ by fixing the parameter $\rho_0 = 0$. We also take $L=1$. The interval $\rho\in(-\infty,0)$ is then mapped to $s\in\left(-1,1\right)$ through 
\begin{eqnarray}
    \label{eq:cheby_map}
    \rho = \log \left(\frac{1+s}{2+s}\right) - \log \left(\frac{2}{3}\right)\ ,
\end{eqnarray}
being this latter adapted to the domain of the Chebyshev polynomials.

The only singular points of the equations of motion for both $Z_\phi$ and $Z_2$ are located at the extrema of the integration domain, $s = \pm 1$. The near-boundary and near-horizon behaviors are
\begin{align}
    Z_\phi &\sim e^{-\frac{2^{1/3}}{3} \im\omega g_\phi(s)}\quad \text{(near horizon)}
    \qquad 
    Z_\phi \sim (1+s)^{3/2}\quad \text{(near boundary)}
    \ , \\
    Z_2 &\sim e^{-\frac{2^{1/3}}{3} \im\omega g_2(s)}\quad \text{(near horizon)}
    \qquad 
    Z_2 \sim (1+s)\quad \text{(near boundary)}\ .
\end{align}
The functions $g_\phi(s)$ and $g_2(s)$ are chosen so as to eliminate the $\omega^2$-term in the equations of motion. This, together with requiring their infalling near-horizon behavior 
\begin{eqnarray}
    g_\phi(s) \sim g_2(s) \sim \log(1-s)\ ,
\end{eqnarray}
determines their derivatives to be 
\begin{equation}
    g_\phi'(s) = g_2'(s)
    = 
    \frac{9\ 2^{2/3} \sqrt[3]{(5 s+7) [s (7 s+16)+13]}}{(s-1) [s (19 s+52)+37]}\ .
\end{equation}

\subsection{Scalar fluctuations: throat inside the horizon}

The numerical approach follows closely what already described in Subsection~\ref{subsec:atthehor}. Now $f_0 > 1$ and, in order to have the horizon located at $\rho = 0$, one fixes
\begin{equation}
    \rho_0
    =
    -\frac{2L}{3}\arctanh\left(\frac{1}{f_0}-1\right)\ ,
\end{equation}
using \eqref{eq:rhoh} and $d=3$.

We map $\rho\to s\in\left(-1,1\right)$ using again the change of variable \eqref{eq:cheby_map}. We find that the only singular points of the equations of motion for both $Z_\phi$ and $Z_2$ are again located at the extrema of the integration domain, $s = \pm 1$. In particular, the near-boundary and near-horizon behaviors are given by
\begin{align}
    Z_\phi &\sim e^{-\frac{1}{3} \left(\frac{11}{6}\right)^{1/3} \im\omega g_\phi(s)}\quad \text{(near horizon)}
    \qquad 
    Z_\phi \sim (1+s)^{3/2}\quad \text{(near boundary)}
    \ , \\
    Z_2 &\sim e^{-\frac{1}{3} \left(\frac{11}{6}\right)^{1/3} \im\omega g_2(s)}\quad \text{(near horizon)}
    \qquad 
    Z_2 \sim (1+s)\quad \text{(near boundary)}\ .
\end{align}
By requiring infalling behavior and that the $\omega^2$-term in the equations of motion is null, one fixes
\begin{equation}
    g_\phi'(s) = g_2'(s)
    = 
    -\frac{54 \sqrt[3]{\frac{5}{11}} \sqrt[3]{27 (s+1)^3+\frac{48}{5} (s+2)^3}}{24 (s+2)^3-81 (s+1)^3}
    \ .
\end{equation}

\subsection{Scalar fluctuations: throat outside the horizon}

As far as the gauge-invariant combination $Z_2$ is concerned, the equation of motion features singular points only at the extrema of the integration domain. The technical aspects of the QNM computations as well as the resulting pattern of modes is qualitatively similar regardless of the relative position of the throat and the horizon.

Regarding $Z_\phi$, instead, when $\frac{1}{2} < f_0 < 1$ and the throat is outside the horizon, a new singular point appears in the integration domain at the throat's position $\rho = \rho_0$, on top of the singular points at the extrema $\rho=-\infty$ and $\rho=\rho_h$. The singular point is related to the factor of $1/A'$ appearing in the definition of $Z_\phi$ in \eqref{eq:fluctuations}. Although $Z_\phi$ develops a pole at the throat where $A'=0$, the metric and scalar fluctuations remain regular. One can make a calculation using the same collocation method with the variable $W_\phi=A' Z_\phi$.

More specifically, the interval $\rho\in(-\infty,0)$ is again mapped to $s\in\left(-1,1\right)$ by means of the radial coordinate change given in \eqref{eq:cheby_map}. The only singular points of the equations for $W_\phi$ and $Z_2$ are located at the extrema of the integration domain, $s = \pm 1$. The near-boundary and near-horizon behaviors are
\begin{align}
    W_\phi &\sim e^{-\mathcal C\,  \im\omega g_\phi(s)}\quad \text{(near horizon)}
    \qquad 
    W_\phi \sim (1+s)^{3/2}\quad \text{(near boundary)}
    \ , \\
    Z_2 &\sim e^{-\mathcal C\,  \im\omega g_2(s)}\quad \text{(near horizon)}
    \qquad 
    Z_2 \sim (1+s)\quad \text{(near boundary)}\ ,
\end{align}
where $\mathcal C$ is a rather complicated expression given in terms of the first root of the equation $16x^3-81=0$; let us just say that it evaluates approximately to $\mathcal C \simeq 0.43679$. The functions $g_\phi(s)$ and $g_2(s)$ are chosen so as to eliminate the $\omega^2$-term in the equations of motion. This, together with requiring their infalling near-horizon behavior 
\begin{eqnarray}
    g_\phi(s) \sim g_2(s) \sim \log(1-s)\ ,
\end{eqnarray}
determines their derivatives to be 
\begin{equation}
    g_\phi'(s) = g_2'(s)
    = 
    \frac{6 \sqrt[3]{3} \sqrt[3]{s [s (167 s+597)+789]+391}}{(s-1) [s (19 s+52)+37]}\ .
\end{equation}

\subsection{Supergravity model}

We used similar methods for the supergravity model, but the calculations were made by Claude Opus 5.5 High. We checked that the code used by Claude and our calculation agree for the fake supergravity models.
After providing the numerical solutions for the backgrounds, the code was adapted to the numerical backgrounds, improved, and run by Claude Opus 5.5 High.

\bibliographystyle{JHEP}
\bibliography{Refs}

\end{document}